\documentclass[pdflatex,sn-mathphys-ay]{sn-jnl}

\usepackage{graphicx}%
\usepackage{multirow}%
\usepackage{array}
\usepackage[table]{xcolor}
\usepackage{amsmath,amssymb,amsfonts}%
\usepackage{caption}
\usepackage{subcaption}
\usepackage{amsthm}%
\usepackage{hyperref}
\usepackage[capitalise]{cleveref}
\crefname{figure}{Figure}{Figures}
\Crefname{figure}{Figure}{Figures} 
\crefname{subsection}{Subsection}{Subsections}
\Crefname{subsection}{Subsection}{Subsections}
\usepackage{mathrsfs}%
\usepackage[title]{appendix}%
\usepackage{xcolor}%
\usepackage{textcomp}%
\usepackage{manyfoot}%
\usepackage{booktabs}%
\usepackage{algorithm}%
\usepackage{algorithmicx}%
\usepackage{algpseudocode}%
\usepackage{listings}%
\usepackage{placeins}   

\theoremstyle{thmstyleone}%
\newtheorem{theorem}{Theorem}
\newtheorem{corollary}{Corollary}

\theoremstyle{thmstyletwo}%
\newtheorem{remark}{Remark}%

\theoremstyle{thmstylethree}%

\begin{document}

\title[Article Title]{Some bivariate distributions on a discrete torus with application to wind direction datasets}


\author[1]{\fnm{Brajesh Kumar} \sur{Dhakad}}\email{brajeshdhakad85047@gmail.com}

\author*[2]{\fnm{Jayant} \sur{Jha}}\email{jayantjha@gmail.com}

\author[3]{\fnm{Debepsita} \sur{Mukherjee}}\email{debepsitamukherjee@gmail.com}

\affil[1]{\orgdiv{Department of Mathematics}, \orgname{Indian Institute of Technology Madras}, \orgaddress{\street{IIT P.O., Sardar Patel Road}, \city{Chennai}, \postcode{600036}, \state{Tamil Nadu}, \country{India}}}

\affil*[2]{\orgdiv{Statistical Science Division}, \orgname{Indian Statistical Institute, Kolkata}, \orgaddress{\street{203 B. T. Road}, \city{Kolkata}, \postcode{700108}, \state{West Bengal}, \country{India}}}

\affil[3]{\orgdiv{Department of Statistics and Data Sciences}, \orgname{University of Texas at Austin}, \orgaddress{\street{ 105 East 24th St. Stop D9800}, \city{Austin}, \postcode{78712}, \state{Texas}, \country{USA}}}


\abstract{Directional measurements such as wind directions are often recorded in a finite number of angular categories rather than as exact angles. When two such measurements are observed jointly, the resulting bivariate observations lie on a discrete torus. Commonly used bivariate circular models are formulated for continuous angular variables. Applying these models to categorical observations requires integrating their densities over regions corresponding to observed category pairs. We propose two parametric models defined directly on the discrete torus, with interpretable parameters for marginal locations and concentrations, and for dependence between the two circular variables. The models provide closed-form probability mass functions and trigonometric moments, which are used to show that, under certain conditions, the dependence parameter characterizes circular--circular correlation. Parameters are estimated by maximum likelihood, and the finite-sample performance is investigated through simulation. The proposed models are applied to three datasets of paired wind direction measurements recorded in 16 equally spaced compass directions at stations in India and compared with discretized versions of established continuous bivariate circular models. They provide competitive fits while allowing likelihood evaluation directly on the observed discrete support. The fitted models are also used to assess the dependence between the paired wind directions in each dataset.
}




\maketitle

\section{Introduction}\label{sec1} 

Directional data arise when observations represent directions, angles, axes, or rotations, with applications in meteorology, environmental science, biology, neuroscience, and computer vision. Comprehensive reviews are provided by \cite{Fisher1995, Jammalamadaka2001, mardia2009, SenGupta2022}. An important setting occurs when two directional measurements are observed jointly, for example, wind directions recorded simultaneously at two locations.

\begin{figure}[!ht]
    \centering
    \includegraphics[width=\textwidth]{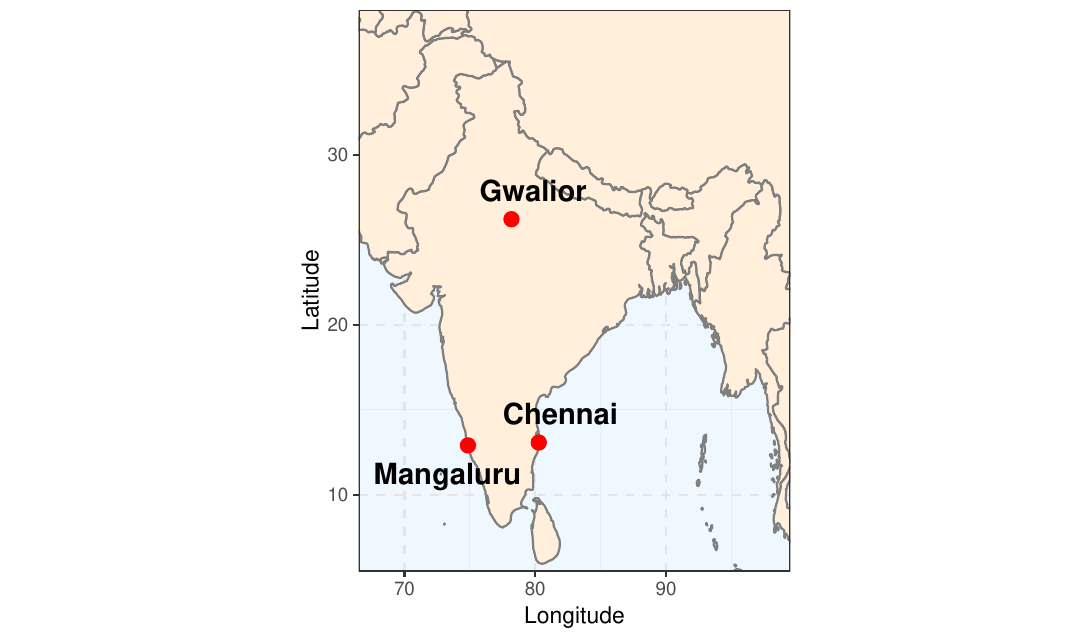}
    \caption{Locations of the three stations used in the wind-direction analysis.}
    \label{fig:1.1}
\end{figure}

Wind direction is an important directional measurement in meteorological and environmental studies, and the relationship between wind directions observed at different locations can provide information about their joint directional behavior. This is particularly interesting in a geographically diverse country such as India, where regional climatic conditions are influenced by distinctive physiographic features, including the Himalayas, the Indian peninsula, and the mountain ranges along different parts of the country \citep{takahashi1981,clift2008,Savindra2006}. Consequently, wind patterns recorded at geographically distinct locations may exhibit different marginal directional characteristics, and investigating whether their directional movements are associated is of scientific interest. In this study, we consider wind direction measurements from Gwalior in north-central India, Chennai on the southeastern coast, and Mangaluru on the southwestern coast; their geographical locations are shown in \cref{fig:1.1}. Our interest is in modeling paired wind directions recorded at these locations and in quantifying how the directions at two stations are related.

A particular feature of these data is that the recorded wind directions are not available as exact angles. Instead, they are reported in 16 equally spaced compass directions, N, NNE, $\ldots$, NNW. Thus, each wind direction observation belongs to a finite set of angular categories. We can index these categories by $k=0,1,\ldots,15$, with the corresponding angular directions represented by $2\pi k/16$. Hence, the circular support is
$$
\left\{\frac{2\pi k}{16}:k = 0,1,\ldots,15 \right\}.
$$
Therefore, when wind directions from two stations are recorded jointly, each observation is of the form
$$
\left(\frac{2\pi k}{16},\frac{2\pi l}{16}\right),\qquad k,l=0,1,\ldots,15,
$$
and the resulting bivariate observations lie on a finite $16\times16$ circular support, which forms a discrete torus.

The scientific question is therefore not only to describe the prevailing wind directions at the individual stations but also to characterize the relationship between the two recorded directions. For example, relative to their respective central directions, the wind directions at two stations may tend to vary in the same rotational sense or in opposite rotational senses around the circle. More generally, the strength and nature of their directional dependence may vary considerably between station pairs. A statistical model for these observations should therefore describe the marginal locations and concentrations of the two wind directions while simultaneously quantifying their dependence on the discrete circular support on which they are actually observed.

More generally, if two directional variables are recorded in $m_1$ and $m_2$ equally spaced angular categories, respectively, their joint support consists of $m_1m_2$ points on the corresponding discrete torus. An important feature of such data is their cyclic rather than linear ordering. Suppose $m$ categories $x_1,x_2,\ldots,x_m$ are arranged around a circle. Then the neighbouring categories of $x_1$ are $x_2$ and $x_m$, whereas under a linear ordering, $x_1$ and $x_m$ lie at opposite ends of the ordering. Statistical models for bivariate discrete circular observations should therefore account simultaneously for the finite support, circular ordering, marginal directional behavior, and dependence between the two variables.

Substantial literature is available for continuously observed bivariate circular variables. Early distributions on the torus were considered by \citet{mardia1975}, while bivariate von Mises models and bivariate wrapped Cauchy models were studied by \citet{mardia2007} and \citet{kato2015}, respectively. Copula-based approaches have also been developed \citep{Jones2015,lagona2024}, and a multimodal bivariate circular distribution was proposed by \citet{hassanzadeh2018}. Circular--circular regression models have also been studied \citep{kato2008,lagona2016,jha2022}. These models provide useful frameworks when the observed variables are continuous angles. They are not, however, probability mass models defined directly on a finite circular support.

A continuous bivariate circular model can be adapted to data recorded in angular categories by associating each observed category pair with a rectangular region on the continuous torus. The probability of an observed category pair is then obtained by integrating the joint density over the corresponding region, and these probabilities are used to construct the likelihood. This provides a useful approach for analyzing such data, but the resulting model is a discretized version of a continuous model rather than a model defined directly on the finite support. When the required probabilities are not available in closed form, their evaluation requires numerical integration.

Discrete circular modeling has been studied more extensively in the univariate setting. Examples include the wrapped Poisson distribution \citep{ball1992}, the wrapped geometric distribution \citep{jacob2013}, and the wrapped symmetric geometric (WSG) distribution \citep{Jha2018}. To the best of our knowledge, the existing discrete circular literature does not provide a parametric distribution on a discrete torus for bivariate discrete circular data. This motivates the development of bivariate probability models defined directly on a discrete torus, with parameters that describe the marginal behavior and dependence between the two circular variables and whose likelihood function can be constructed directly from the probability masses of the observed category pairs.

To address this problem, we propose two parametric distributions defined directly on a discrete torus: the bivariate wrapped geometric (BWG) distribution and its generalized form, the bivariate generalized wrapped geometric (BGWG) distribution. The models are constructed using WSG distributions together with a trigonometric term that introduces dependence between the two circular variables. Their parameters characterize the marginal locations and concentrations as well as the dependence between the two recorded directions. In particular, the dependence structure allows different patterns of directional association, including rotational and anti-rotational dependence.

The proposed models have closed-form probability mass functions and trigonometric moments. Consequently, the probability of each observed category pair is obtained directly from the joint pmf, without requiring the integration used to discretize continuous bivariate circular models. The closed-form trigonometric moments are used to establish the relationship between the model's association structure and a circular--circular correlation coefficient. These features provide an interpretable framework for describing both the marginal wind-direction patterns at individual stations and the association between directions recorded at different stations.

Inference is carried out by maximum likelihood. Because both models contain continuous and discrete parameters, estimation is performed by considering all possible values of the discrete parameters and numerically maximizing the likelihood over the continuous parameters. The estimates corresponding to the largest maximized likelihood are then selected as the maximum likelihood estimates. The finite-sample behavior of the resulting estimators is examined through Monte Carlo simulation under different sample sizes and parameter configurations.

The proposed models are further applied to three bivariate wind direction datasets formed from the station pairs Gwalior--Mangaluru, Chennai--Gwalior, and Chennai--Mangaluru. For comparison, established continuous bivariate circular models, namely the bivariate wrapped Cauchy distribution \citep{kato2015} and the bivariate von Mises Sine and Cosine distributions \citep{mardia2007}, are adapted to the same discrete observations by integrating their densities over the corresponding regions on the continuous torus. Across the three datasets, the proposed models provide competitive fits while modeling the observed category pairs directly on the discrete torus.
 
This article is organized as follows: \cref{sec2} introduces the BWG distribution for bivariate discrete circular observations, and \cref{sec3} examines its distributional properties. \cref{sec4} develops the BGWG distribution, while \cref{sec5} investigates the dependence parameter and its relationship with a circular--circular correlation measure. Maximum likelihood estimation is described in \cref{sec6}, and the finite-sample performance of the estimators is studied through simulation in \cref{sec7}. \cref{sec8} applies the proposed models to three wind direction datasets and compares their fits with discretized continuous bivariate circular models. \cref{sec9} concludes the article.    

\section{Bivariate model for discrete circular data}\label{sec2}
We briefly discuss the WSG distribution in \cref{subsec.2.1} for univariate ordinal circular data. Then, in \cref{subsec.2.2}, we construct a bivariate distribution based on WSG and a cosine function.  

\subsection{Wrapped symmetric geometric distribution}\label{subsec.2.1} \vspace{-0.5 mm}
Let $\mathbb{Z}_+$ denote the set of positive integers, $\mathbf{Z}_m$ the cyclic group of integers of order $m$, and $\mathbf{D}_m$ the domain consisting of the vertices of a regular polygon, i.e.,
\begin{equation*}
\scalebox{0.85}{$
\begin{aligned}
     \mathbf{Z_m} = \{0,1,\cdots,m-1\}, \quad\text{and}\quad \mathbf{D_m} = \Big\{\frac{2\pi r}{m}, r \in \mathbf{Z_m} \Big\}, \quad\text{where}\quad m \in \mathbb{Z_+}. 
\end{aligned}
 $}  
\end{equation*}
The probability mass function (pmf) of a discrete random variable is denoted as $P(\cdot)$. The pmf of the discrete random variable $X$, defined on $\mathbf{D_m}$, is given as
\begin{equation}
\label{eq:2.2}
\resizebox{0.6\textwidth}{!}{$
P\!\left(\frac{2\pi k}{m}\right)
=
\frac{(1-q)\left(q^{\zeta}+q^{m-\zeta}\right)}
{(1-q^m)(1+q)},
\qquad k\in\mathbf{Z}_m.
$}
\end{equation}
where $q \in (0,1)$, $\zeta = (k-\alpha) \bmod m$, and $\alpha \in \mathbf{Z_m}$.
The distribution given in \eqref{eq:2.2} is called WSG distribution for ordinal circular data and is explored by \cite{Jha2018}. It is a unimodal symmetric distribution whose modal direction is ${\frac{2 \pi \alpha}{m}}$. 
The WSG distribution is undefined for $ q \in \{0,1\} $; these cases are examined by evaluating the limit of the pmf \eqref{eq:2.2}. When $ q = 0 $, it degenerates at $ \frac{2\pi\alpha}{m} $, whereas for $ q = 1 $, it becomes uniform. 

The study of bivariate wind directions motivates the need for a generalization of the WSG distribution. In \cref{subsec.2.2}, we present a bivariate distribution on the discrete torus whose sample space is $\mathbf{D_{m_1}}\times \mathbf{D_{m_2}} = \Big\{\frac{2\pi k}{m_1}, k \in \mathbf{Z_{m_1}} \Big\} \times \Big\{\frac{2\pi l}{m_2}, l \in \mathbf{Z_{m_2}}\Big\}$, which represents $m_1\times m_2$ grid points on the torus.

\subsection{Bivariate wrapped geometric distribution}\label{subsec.2.2} 
The random variable $({X_1}, {X_2})$ is a bivariate circular random variable defined on $\mathbf{D_{m_1}}\times \mathbf{D_{m_2}}$, and its pmf is denoted by $P(\cdot, \cdot)$. It can be written as
\begin{equation*}
\resizebox{0.65\textwidth}{!}{$
\begin{aligned}
 P\bigg( \frac{2\pi k}{m_1},\frac{2\pi l}{m_2}\bigg) = P\bigg({X_1} = \frac{2\pi k}{m_1}, {X_2} = \frac{2\pi l}{m_2}\bigg),~k\in \mathbf{Z_{m_1}}, \vspace{0.3mm} l \in \mathbf{Z_{m_2}}.   
\end{aligned}
$}
\end{equation*}
The proposed joint distribution of $({X_1}, {X_2})$ is given by:
\begin{equation}
\label{eq:2.3}
\scalebox{0.98}{$
P\!\left( \frac{2\pi k}{m_1},\frac{2\pi l}{m_2}\right) \propto\left(q^{\zeta_1}+ q^{m_1-\zeta_1}\right)\left(s^{\zeta_2} + s^{{m_2}-\zeta_2}\right)\left(1 + \rho \cos\left(\frac{2\pi \zeta_1}{m_1} - \delta \frac{2\pi \zeta_2}{m_2}\right)\right),
$}
\end{equation}
where $\zeta_1 = (k-\alpha)\bmod {m_1}$ and $\zeta_2 = (l-\beta)\bmod {m_2}$. The parameters are $\alpha \in \mathbf{Z}_{m_1},~\beta \in \mathbf{Z}_{m_2},~\rho \in [-1,1],~\delta \in \{-1, 1\}$ and $q,s \in (0,1)$.

The proposed distribution has a copula interpretation similar to that of the circula described by \cite{Jones2015}, where the circula employs the discrete version of cardioid density defined by  
\begin{equation*}
\scalebox{0.85}{$
\begin{aligned}
\mathrm{g}\! \left(\frac{2\pi k}{m }\right) = \frac{1}{m} \left( 1 + \rho \cos\left(\frac{2\pi \zeta}{m}\right) \right).
\end{aligned}
$}
\end{equation*}
Next, we derive a few theorems and corollaries related to the proposed distribution. All proofs are given in Appendix. 
\begin{theorem}\vspace{-2mm}
\label{theo.2.1}
The closed form expression for the proposed joint pmf of $(X_1, X_2)$ is given by
\begin{equation}
\label{eq:2.5}
\resizebox{0.95\textwidth}{!}{$
P\!\left(\frac{2\pi k}{m_1},\frac{2\pi l}{m_2}\right) = C_1\left(q^{\zeta_1}+q^{m_1-\zeta_1}\right)\left(s^{\zeta_2}+s^{m_2-\zeta_2}\right)\left(1+\rho\cos\left(\frac{2\pi\zeta_1}{m_1}-\delta\frac{2\pi\zeta_2}{m_2}\right)\right), 
$}
\end{equation}
where the normalizing constant is given by 
\[\begin{aligned} C_1& = \frac{(1-q)(1-s)}{(1-q^{m_1})(1-s^{m_2})(1+q)(1+s)}\\
&\quad \times \frac{\left(q^2-2q\cos\frac{2\pi}{m_1}+1\right)\left(s^2 -2s\cos\frac{2\pi}{m_2} + 1\right)}{\left(q^2-2q\cos\frac{2\pi}{m_1}+1\right)\left(s^2 -2s\cos\frac{2\pi}{m_2} + 1\right) + \rho(1-q)^2(1-s)^2}. \end{aligned}
\]
\end{theorem} 
\noindent The proof is given in \autoref{app1}.

The distribution given in \eqref{eq:2.5}, referred to as ``bivariate wrapped geometric'' (BWG) distribution, is a discrete bivariate circular distribution symmetric around $\left(\frac{2 \pi \alpha}{m_1},\frac{2\pi \beta}{m_2}\right)$. Since the BWG distribution is not defined when parameters $q$ or $s$ take values in $\{0,1\}$, the distributions for these cases are derived by evaluating the limits of the pmf \eqref{eq:2.5}. The results are summarized in \cref{tab:2.1}.

\section{Properties of the model}\label{sec3} 
In this section, we discuss some properties of the BWG distribution, including the role of parameters in the dependence structure of $(X_1, X_2)$, the dominant mode (i.e., the point of highest probability), and the marginal and conditional distributions. 

\subsection{Dependence Structure and Modality}
We examine the role of the parameter $\rho$ in the proposed BWG distribution. We first establish the condition under which the circular random variables $X_1$ and $X_2$ are independent and then investigate the effect of $\rho$ on the modality of the joint pmf.

\begin{corollary}
\label{coro.3.1}
In the BWG distribution, the random variables $X_1$ and $X_2$ are independent if and only if $\rho=0$.
\end{corollary} 
\noindent The proof is given in \autoref{app2}.

\begin{figure}[!ht]
\centering
\centering
\includegraphics[width=\linewidth]{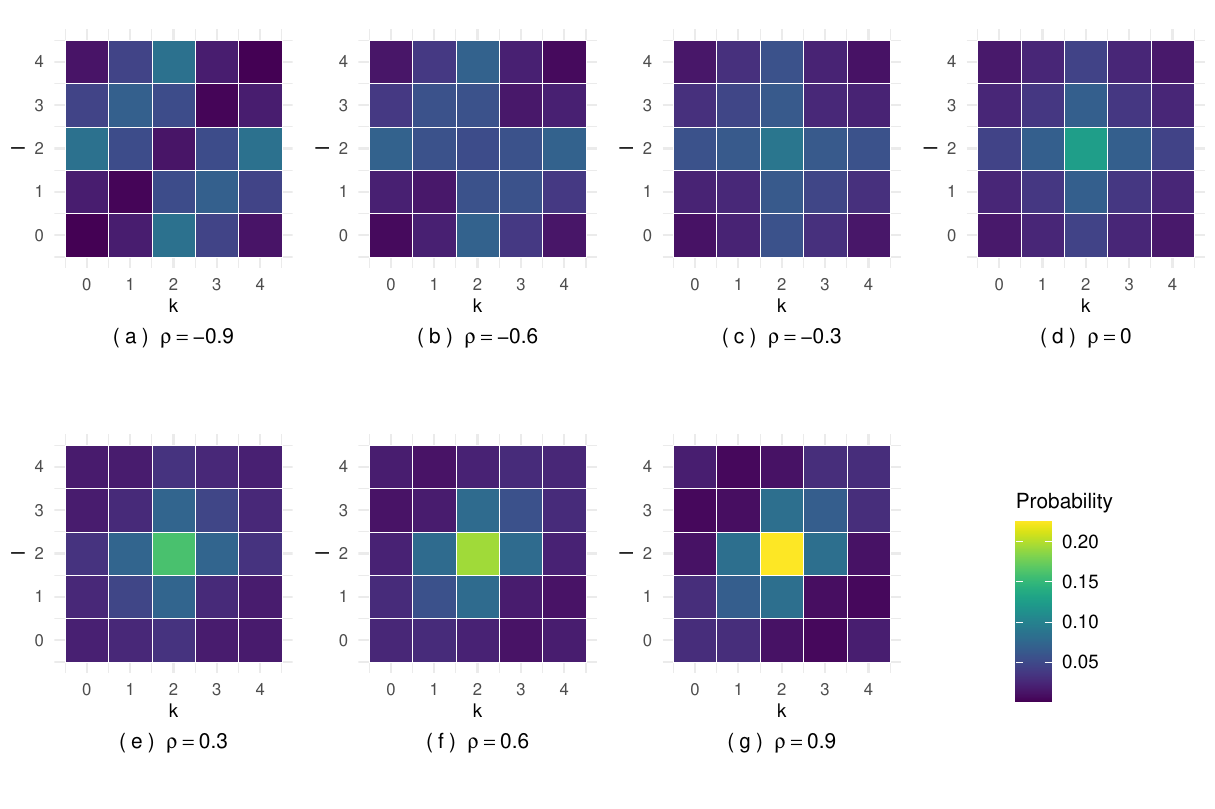}
\caption{Probability mass plot of BWG distribution for various values of $\rho$}\vspace{-2mm}
\label{fig:3.1}
\end{figure}

Corollary~\ref{coro.3.1} motivates a graphical investigation of the effect of $\rho$ on the joint pmf. To illustrate this, we fix $m_1=m_2=5$, $q=s=0.5$, $\alpha=\beta=2$, and $\delta=1$. Figure~\ref{fig:3.1} presents the probability mass plots of the BWG distribution for various values of $\rho$.

In \cref{fig:3.1}, the cell $(k,l)$ corresponds to the location $\left(\frac{2 \pi k}{m_1},\frac{2 \pi l}{m_2}\right)$. Subfigures (a) and (b) illustrate that the dominant mode is not unique. Specifically, the maximum probability is attained at
$\left(0,\frac{4\pi}{5}\right)$,
$\left(\frac{8\pi}{5},\frac{4\pi}{5}\right)$,
$\left(\frac{4\pi}{5},0\right)$, and
$\left(\frac{4\pi}{5},\frac{8\pi}{5}\right)$,
which correspond to the cells $(0,2)$, $(4,2)$, $(2,0)$, and $(2,4)$, respectively. Under the circular ordering, the cells $(0,2)$ and $(4,2)$ are neighbours, and the cells $(2,0)$ and $(2,4)$ are also neighbours. In contrast, subfigures (c) to (g) illustrate that the dominant mode is unique at $\left(\frac{2 \pi \alpha}{m_1},\frac{2 \pi \beta}{m_2}\right)$. The above observations indicate that the modality of the proposed BWG distribution varies with $\rho$. The following theorem establishes the uniqueness of the dominant mode for $\rho \geq 0$.

\begin{theorem}
\label{theo.3.2}
For $\rho \geq 0$, the BWG distribution has a unique dominant mode at $\left(\frac{2\pi\alpha}{m_1},\frac{2\pi\beta}{m_2}\right)$.
\end{theorem} 
\noindent The proof is given in \autoref{app3}.

Theorem~2 establishes the location and uniqueness of the dominant mode of the BWG distribution for $\rho\ge0$. Evaluating the limiting form of the pmf \eqref{eq:2.5} at the boundary point $q=s=1$ shows that the resulting distribution behaves differently from that described in Theorem~2. Specifically, if $\rho>0$, the maximum probability is attained at all points satisfying
\[
\cos\left(\frac{2\pi\zeta_1}{m_1}-\delta\frac{2\pi\zeta_2}{m_2}\right)=1,
\]
whereas if $\rho=0$, the resulting distribution is uniform. This motivates an investigation of the behavior of the distribution at the boundary values of the parameters $q$ and $s$.  Table~\ref{tab:2.1} summarizes the corresponding limiting distributions.

\begin{table}[!ht]
\centering
\caption{Limits of pmf \eqref{eq:2.5} on boundaries of parameter space}\vspace{-2mm}
\label{tab:2.1}
\begin{tabular}{cc}
 \hline
 $(q,s)$ & pmf $P(\cdot, \cdot)$\\
 \hline
 & \\
 $\{0\}\times(0, 1)$ & 
 $\begin{cases}
 C_2(s^{\zeta_2} + s^{m_2-\zeta_2})\big(1 +\rho \cos\frac{2 \pi \zeta_2}{m_2}\big) &\hspace{0.1 cm}\text{if}\hspace{0.1 cm} k =\alpha\text{,}\hspace{0.15 cm}l \in \mathbf{Z_{m_2}}\\ 
 0 &\hspace{0.1cm}\text{else},
 \end{cases}$ \\
 
 &where $ C_2 = \Big(\frac{1}{(1-s^{m_2})(1+s)}\Big)\bigg(\frac{(1-s)(s^2 -2s\cos\frac{2\pi}{m_2} +1)}{s^2 -2s\cos\frac{2\pi}{m_2} +1 +\rho(1-s)^2}\bigg)$\\
 &\\
 $(0, 1)\times\{0\}$ &
 $\begin{cases}
 C_3\big(q^{\zeta_1} + q^{m_1 -\zeta_1}\big)\big(1 +\rho \cos\frac{2 \pi \zeta_1}{m_1}\big) & \hspace{0.2cm} \text{if}\hspace{0.1cm} k \in \mathbf{Z_{m_1}}\text{,} \hspace{0.15cm} l = \beta\\
 0 & \hspace{0.2cm}\text{else,}
 \end{cases}$\\
 
 & where $ C_3  = \Big(\frac{1}{(1 -q^{m_1}) (1+q)}\Big)\bigg(\frac{(1 -q)\big(q^2 -2q \cos\frac{2\pi}{m_1} + 1\big)}{q^2 -2q \cos\frac{2\pi}{m_1} + 1 + \rho (1-q)^2} \bigg) $ \\
 & \\
 $\{0\}\times\{0\}$ & $\begin{cases}
 1 &\hspace{0.2 cm}\text{if}\hspace{0.2 cm} k =\alpha\text{,}\hspace{0.15 cm}l = \beta\\
 0 &\hspace{0.2cm}\text{else}
 \end{cases} $ \\
 
   & \\   
 $\{1\}\times(0, 1)$ & $ \frac{1}{m_1}\bigg(\frac{(1-s)\big(s^{\zeta_2} +s^{m_2-\zeta_2}\big)}{(1-s^{m_2})(1+s)}\bigg)\bigg(1 + \rho \cos\Big(\frac{2\pi \zeta_1}{m_1} - \delta \frac{2\pi \zeta_2}{m_2}\Big) \bigg) $ \\
 & \\
 $(0, 1)\times\{1\}$ & $\frac{1}{m_2}\bigg(\frac{(1-q)\big(q^{\zeta_1} + q^{m_1-\zeta_1}\big)}{(1-q^{m_1})(1 +q)}\bigg)\bigg(1 + \rho \cos\Big(\frac{2\pi \zeta_1}{m_1} - \delta \frac{2\pi \zeta_2}{m_2}\Big)\bigg)$\\
 & \\
 $\{1\}\times\{1\}$ & $\frac{1}{{m_1}\times{m_2}}\bigg(1 + \rho \cos\Big(\frac{2\pi \zeta_1}{m_1} - \delta \frac{2\pi \zeta_2}{m_2}\Big) \bigg)$\\
 & \\
 $\{1\}\times\{0\}$ & $\begin{cases}
 \frac{1}{m_1}\big(1 + \rho \cos\frac{2\pi \zeta_1}{m_1}\big)&\quad \text{if}\hspace{0.1cm} k \in \mathbf{Z_{m_1}}\text{,} \hspace{0.15cm} l = \beta\\
 0 & \quad \text{else} 
 \end{cases} $\\
 & \\
 $\{0\}\times\{1\}$ & $\begin{cases}
 \frac{1}{m_2}\big(1 +\rho \cos\frac{2 \pi \zeta_2}{m_2}\big) &\hspace{0.2 cm}\text{if}\hspace{0.2 cm} k =\alpha\text{,}\hspace{0.15 cm} l \in \mathbf{Z_{m_2}}\\
 0 &\hspace{0.2cm}\text{else}
 \end{cases} $ \\
 \hline 
\end{tabular}
\end{table}

\subsection{Marginal wrapped symmetric geometric distribution}
The marginal pmf of $X_1$ is given by
\begin{equation}
\label{eq:3.1}
\scalebox{0.85}{$\begin{aligned}
 P\!\left(\frac{2\pi k}{m_1}\right)& = \sum_{l=0}^{m_2-1}P\!\left(\frac{2\pi k}{m_1}, \frac{2\pi l}{m_2} \right)\\
 & = C_4\left(q^{\zeta_1}+q^{m_1-\zeta_1}\right)\left(s^2-2s\cos\frac{2\pi}{m_2}+1 +\rho\left(1-s\right)^2\cos\frac{2\pi \zeta_1}{m_1}\right),
\end{aligned}
$}
\end{equation} 
where 
\[ 
 C_4 = \frac{1-q}{(1-q^{m_1})(1+q)} \frac{q^2-2q\cos\frac{2\pi}{m_1}+1}{\left(q^2 -2q\cos\frac{2\pi}{m_1}+1\right)\left(s^2-2s\cos\frac{2\pi}{m_2}+1\right)+\rho(1-q)^2(1-s)^2}. 
 \]

The marginal distribution in \eqref{eq:3.1} is symmetric about the direction
$\frac{2\pi\alpha}{m_1}$ and is referred to as the ``marginal wrapped symmetric geometric'' (MWSG) distribution of $X_1$. The marginal distribution of $X_2$ can be obtained analogously and is symmetric about the direction $\frac{2\pi\beta}{m_2}$.

\begin{theorem}
\label{theo.3.3}
For $\rho\geq0$, excluding the limiting cases where $q=s=1$ or $q=1$ and $\rho=0$, the MWSG distribution is unimodal with mode at $\frac{2\pi \alpha}{m_1}$.   
\end{theorem} 
\noindent The proof is given in \autoref{app4}.

This implies that for nonnegative $\rho$, the distribution is unimodal, except in the specific limiting cases where it reduces to the uniform distribution. These limiting cases are summarized in \cref{tab:3.1}.

\begin{table}[!ht]
\centering
\caption{Limits of pmf \eqref{eq:3.1} on boundaries of parameter space}
\label{tab:3.1}
\begin{tabular}{cc}
\hline
 $(q,s)$ & pmf $P(\cdot)$ \\
\hline
 $(0, 1)\times\{1\}$ & $\Big(\frac{1 -q}{(1 -q^{m_1})(1 +q)}\Big)\big(q^{\zeta_1} +q^{m_1-\zeta_1}\big)$\\ 
 &\\
 $\{ 1\}\times(0, 1)$ & $\frac{1}{m_1}\bigg(1 + \rho \bigg(\frac{(1 -s)^2 \cos\frac{2 \pi \zeta_1}{m_1}}{s^2 -2s \cos\frac{2 \pi}{m_2} +1 }\bigg)\bigg)$\\
 &\\
 $(0, 1)\times\{0\}$ & $C_5\big(q^{\zeta_1} +q^{m_1-\zeta_1}\big)\big(1 +\rho \cos\frac{2\pi \zeta_1}{m_1}\big),$ \\
 & \\
 & where $C_5=\bigg(\frac{(1-q)(q^2 -2q\cos\frac{2\pi}{m_1} +1)}{(1-q^{m_1})(1+q)\big((q^2 -2q\cos\frac{2\pi}{m_1} +1) + \rho(1-q)^2\big)}\bigg)$\\
 & \\
 $\{ 0\}\times(0, 1)$ & $\begin{cases}
 1 \hspace{.5 cm} \text{if} \hspace{0.3 cm} k = \alpha\\
 0 \hspace{0.5 cm} \text{else} 
 \end{cases}$\\
 & \\
 $\{ 1\}\times\{ 1\}$ & $\frac{1}{m_1}, \hspace{0.5 cm}\text{ for}\hspace{0.3 cm} k\in\mathbf{Z_{m_1}}$\\
 & \\
\hline
\end{tabular}  
\end{table}

\cref{tab:3.1} summarizes the limiting forms of the pmf on the boundaries of the parameter space of $(q,s)$. It highlights how specific parameter combinations lead to simplified distributions, such as degenerate or uniform cases, consistent with the conditions discussed in \cref{theo.3.3}. The first case in this table, corresponding to  $(q, s) \in (0, 1)\times\{1\}$, is the WSG distribution.

\subsection{Conditional wrapped geometric distribution}
Here, we derive the conditional distribution of $X_2$ given $X_1 =\frac{2\pi k}{m_1}$, expressed as
{\small
\begin{equation}
\label{eq:3.3}
\begin{aligned}
P\!\left(\frac{2\pi l}{m_2}\bigg|\frac{2 \pi k}{m_1}\right)& = 
C_6\left(\frac{\left(s^{\zeta_2} + s^{m_2-\zeta_2}\right)\left(1 + \rho \cos\left(\frac{2\pi \zeta_1}{m_1} - \delta \frac{2\pi \zeta_2}{m_2}\right)\right)}{1 -2s\cos\frac{2\pi}{m_2} + s^2+ \rho (1-s)^2\cos\frac{2\pi \zeta_1}{m_1}}\right) ,
\end{aligned}
\end{equation}
}
 where, {\small $ C_6 = \Big(\frac{1-s}{(1-s^{m_2})(1+s)}\Big)\big(1 -2s\cos\frac{2\pi}{m_2} + s^2\big) $}. 

\begin{figure}[!ht]
  \centering
  \centering
  \includegraphics[width= \linewidth, height=0.45\textheight]{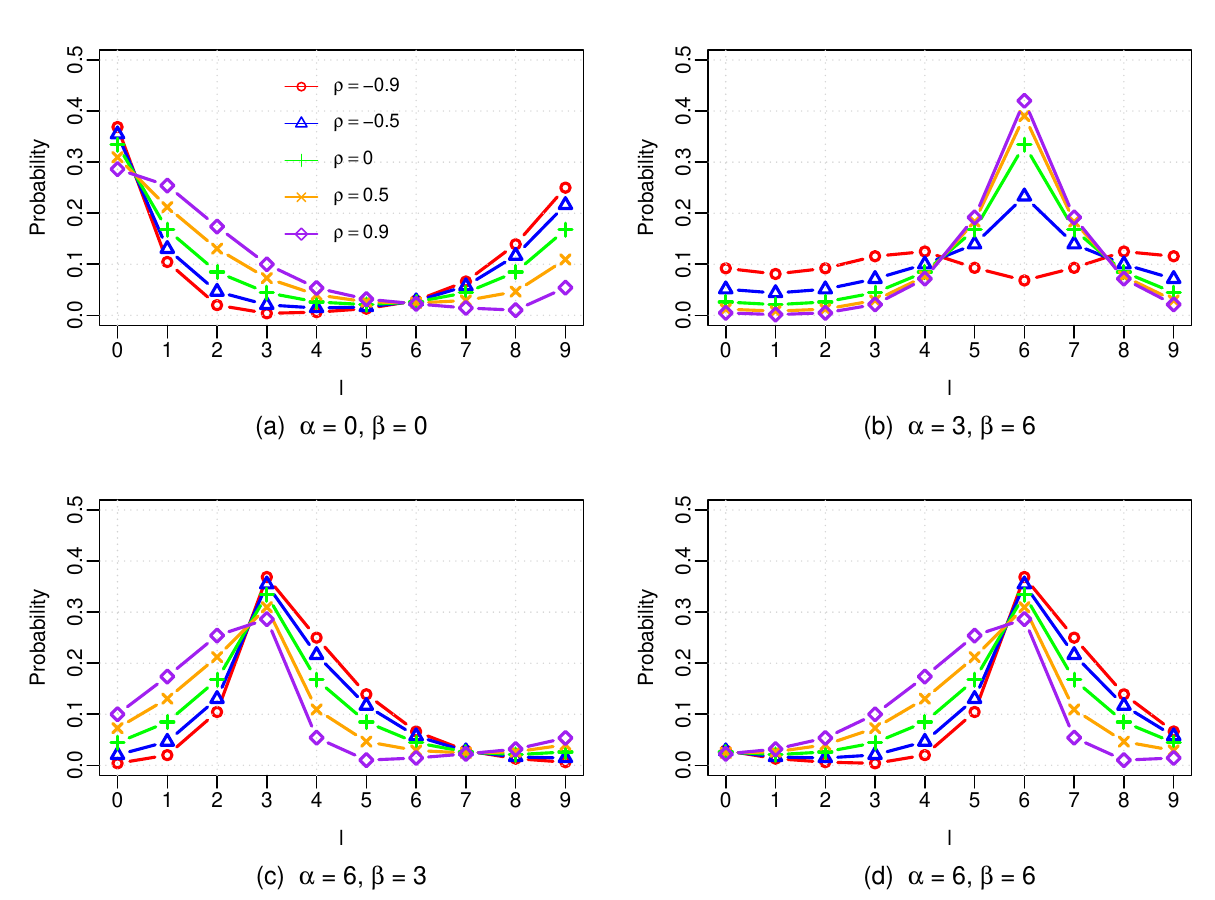}   
 \caption{Probability mass plot of CWG distribution for fixed $X_1 = 6 \pi/m_1$}\vspace{-2mm}
 \label{fig:3.2}
\end{figure}

The distribution given in \eqref{eq:3.3} is referred to as the ``conditional wrapped geometric'' (CWG) distribution. \cref{fig:3.2} shows how the parameters $\alpha$, $\beta$, and $\rho$ affect the conditional distribution of $X_2$, under the settings $m_1 = m_2 = 10,~q = s = 0.5$, and $\delta = 1$, given that $X_1 = 6\pi/ m_1$.

In \cref{fig:3.2}, the case $\rho = 0$ represents the marginal pmf of $X_2$, which is the WSG distribution, since $X_1$ and $X_2$ are independent. As evident in Subfigures (a), (c), and (d), as the magnitude of $\rho$ increases, the conditional distribution becomes asymmetric, reflecting the emergence of dependence. Generally, the dominant mode remains at $\frac{2\pi \beta}{m_2}$. However, for specific settings such as $\rho = -0.9$, $\alpha = 3$, and $\beta = 6$ (see subfigure (b)), it shifts away from this direction, illustrating how the parameters $\alpha$, $\beta$, and $\rho$ jointly influence the shape and asymmetry of the conditional distribution.

\section{Generalized BWG distribution}\label{sec4}
In this section, we generalize the pmf \eqref{eq:2.5} by allowing $\alpha \in [0, m_1)$ and $\beta \in [0, m_2)$. This generalization introduces asymmetry into the distribution, except when $\alpha$ and $\beta$ are integers. The generalized form of this pmf is expressed as

\begin{align}
\label{eq:4.1}
 P\!\left( \frac{2\pi k}{m_1},\frac{2\pi l}{m_2}\right) = C_7 \left(q^{\zeta_1} + q^{m_1-\zeta_1}\right)\left(s^{\zeta_2} + s^{m_2-\zeta_2}\right)\left(1 +\rho \cos\left(\frac{2\pi \zeta_1}{m_1} - \delta \frac{2\pi \zeta_2}{m_2}\right)\right),
\end{align}
where $C_7$ is the normalizing constant. Its reciprocal is given by

\begin{align*}
 \frac{1}{C_7} &= \sum_{n_1 = 0}^{m_1-1}\sum_{n_2 = 0}^{m_2-1}\left(q^{n_1+a} + q^{m_1 - n_1 -a}\right)\left(s^{n_2 + b} +s^{m_2 - n_2 -b}\right)\bigg(1\\&\quad +\rho \cos\left(\frac{2\pi(n_1 +a)}{m_1} - \delta \frac{2\pi(n_2 +b)}{m_2}\right)\bigg)\\
 &=\left(1-q^{m_1}\right)\left(1-s^{m_2}\right)\Bigg(\frac{\big(q^{a} +q^{1-a}\big)\left(s^b + s^{1-b}\right)}{(1-q)(1-s)} \\
 &\quad +\rho\frac{C_8}{\left(q^2 -2q\cos\frac{2\pi}{m_1} +1\right)\left(s^2 -2s\cos\frac{2\pi}{m_2} +1\right)}\Bigg),
\end{align*}
where
\begin{align*}
 C_8 &= \left(s^b-s^{2-b}\right)\Bigg(\left(q^a -q^{2-a}\right)\cos\left(\frac{2\pi a}{m_1} - \delta \frac{2\pi b}{m_2}\right)\\&\quad+\left(q^{1-a}-q^{1+a}\right)\cos\bigg(\frac{2\pi(1-a)}{m_1} +\delta \frac{2\pi b}{m_2}\bigg)\Bigg)\\& \quad+\left(s^{1-b}-s^{1+b}\right)\Bigg(\left(q^{a}-q^{2-a}\right)\cos\left(\frac{2\pi a}{m_1} + \delta \frac{2\pi(1-b)}{m_2}\right)\\& \quad+\left(q^{1-a} - q^{1+a}\right)\cos\left(\frac{2\pi(1-a)}{m_1} - \delta \frac{2\pi (1-b)}{m_2}\right)\Bigg) .
\end{align*}

The distribution given in \eqref{eq:4.1} is referred to as the ``bivariate generalized wrapped geometric'' (BGWG) distribution.
The quantities $a$ and $b$ appearing in the normalizing constant $C_7$ represent the fractional parts of $\alpha$ and $\beta$, respectively.

\begin{corollary}
\label{coro.4.1}
In the BGWG distribution, the random variables $X_1$ and $X_2$ are independent if and only if $\rho=0$.
\end{corollary}
\noindent The proof is given in \autoref{app5}.

This result mirrors the independence property of the BWG distribution, confirming that the generalization preserves this structural characteristic.

\section{Measuring dependence using \texorpdfstring{$\rho$}{rho}}\label{sec5} 
 
Since \cref{coro.3.1} and \cref{coro.4.1} establish that the random variables $X_1$ and $X_2$ are independent when $\rho = 0$, this motivates an examination of the role of $\rho$ in their dependence structure. In a bivariate directional framework, interdependence between circular random variables is typically quantified through a correlation coefficient. Accordingly, we derive a relationship between $\rho$ and an appropriate correlation coefficient for the circular random variables $X_1$ and $X_2$.
 
\cite{jupp1980} proposed a correlation coefficient for the bivariate circular case by embedding the circular variables $X_1$ and $X_2$ into the Euclidean plane. Specifically, each variable is defined as $t_1'(X_1) = (\cos X_1, \sin X_1)$ and $t_2'(X_2) = (\cos X_2, \sin X_2)$. The covariance matrix of the joint vector $(t_1'(X_1), t_2'(X_2))$ is given by

$$
\Sigma = \begin{bmatrix} \Sigma_{11} & \Sigma_{12} \\ \Sigma_{21} & \Sigma_{22} \end{bmatrix},
$$
and the correlation coefficient is defined as 
$
\rho^2_1 = \text{tr}(\Sigma_{11}^{-1} \Sigma_{12} \Sigma_{22}^{-1} \Sigma_{21}). 
$
An explicit expression for $\rho_1^2$ is given by

\begin{align*}
\rho^2_1& = \{\rho^2_{1cc} + \rho^2_{1cs} + \rho^2_{1sc} + \rho^2_{1ss} + 2(\rho_{1cc}\rho_{1ss} + \rho_{1cs}\rho_{1sc})\rho'_1\rho'_2 - 2(\rho_{1cc}\rho_{1sc}\nonumber\\
&\quad\quad+ \rho_{1cs}\rho_{1ss})\rho'_1 - 2(\rho_{1cc}\rho_{1cs} + \rho_{1sc}\rho_{1ss})\rho'_2\}/\{(1 - \rho'^2_1)(1 - \rho'^2_2)\},   
\end{align*}
where $\rho_{1cc}= \operatorname{corr}(\cos X_1, \cos X_2),~ \rho_{1cs} = \operatorname{corr}(\cos X_1, \sin X_2)$, etc., and $\rho'_1 = \operatorname{corr}(\cos X_1, \sin X_1)$, $\rho'_2 = \operatorname{corr}(\cos X_2, \sin X_2)$, and $\operatorname{corr}(\cdot,\cdot)$ denotes the usual correlation coefficient.

\begin{theorem}\label{theo.5.1}
For the BWG distribution, if $\rho > 0$, then the correlation coefficient $\rho^2_1$ is monotonically increasing in $\rho$, and if $\rho < 0$, then it is monotonically decreasing in $\rho$. 
\end{theorem}
\noindent The proof is given in \autoref{app6}.

\begin{remark}\label{rem.5.1}
For the BGWG distribution, if $\rho > 0$, then the correlation coefficient $\rho^2_1$ is monotonically increasing in $\rho$, and if $\rho < 0$, then it is monotonically decreasing in $\rho$. (This property is numerically verified for $m_1, m_2 \leq 600$, details are provided in the supplementary material.)    
\end{remark} 
These results indicate that the parameter $\rho$ governs the strength of dependence between $X_1$ and $X_2$. Larger absolute values of $\rho$ correspond to stronger association. The interpretation of $\rho$ is discussed in the following subsection.

\subsection{Interpretation of \texorpdfstring{\(\rho\)}{rho} in BWG and BGWG distributions }

According to \cref{theo.5.1} and \cref{rem.5.1}, for the BWG and BGWG distributions, the parameter $\rho$ is directly related to the correlation coefficient $\rho_1^2$.

For $\rho > 0$, the type of dependence is determined by $\delta$:
\begin{itemize}
    \item If $\delta = 1$, $X_1$ and $X_2$ exhibit a rotational dependence, meaning they move in the same direction, either clockwise or anticlockwise, relative to their mean directions together.
    \item If $\delta = -1$, $X_1$ and $X_2$ exhibit an anti-rotational dependence, meaning they move in the opposite directions, one clockwise and another anticlockwise, relative to their mean directions together.
\end{itemize}
 For $\rho = 0$, no rotational dependence exists; in fact, they are independent. When $\rho < 0$, the variables exhibit dependence, but this cannot be explicitly stated in terms of rotational or anti-rotational dependence.

\section{Parameter estimation}\label{sec6}

Suppose that $\left\{\left(\frac{2\pi k_i}{m_1},\frac{2\pi l_i}{m_2}\right):i=1,\ldots,n\right\}$, where $\ k_i\in \mathbf{Z_{m_1}}$ and $l_i\in \mathbf{Z_{m_2}}$, is a random sample from the BWG distribution. Let
\[\boldsymbol{\theta} = (q,s,\rho,\alpha,\beta,\delta)^\top \] 
denote the vector of unknown model parameters, where $q,s\in(0,1)$, $\rho\in[-1,1]$, $\alpha\in \mathbf{Z_{m_1}}$, $\beta\in \mathbf{Z_{m_2}}$, and $\delta\in\{-1,1\}$. The likelihood function is given by   
\[
L(\boldsymbol{\theta})
= \prod_{i=1}^{n}C_1\left(q^{\zeta_{1_i}}+q^{m_1-\zeta_{1_i}}\right)\left(s^{\zeta_{2_i}}+s^{m_2-\zeta_{2_i}}\right)\left(1+\rho\cos\left(\frac{2\pi\zeta_{1_i}}{m_1}-\delta\frac{2\pi\zeta_{2_i}}{m_2}\right)\right),
\]
where $\zeta_{1_i} = (k_i-\alpha)\bmod {m_1}$ and $\zeta_{2_i} = (l_i-\beta)\bmod {m_2}$. The log--likelihood function is
\begin{align}\label{eq:6.1}
\ell(\boldsymbol{\theta}) = &n\log C_1 + \sum_{i=1}^{n} \bigg(\log \left(q^{\zeta_{1_i}}+q^{m_1-\zeta_{1_i}}\right)
+ \log \left(s^{\zeta_{2_i}}+s^{m_2-\zeta_{2_i}}\right) \nonumber\\ +& \log \left(1+\rho\cos\left(\frac{2\pi\zeta_{1_i}}{m_1}-\delta\frac{2\pi\zeta_{2_i}}{m_2}\right)\right) \bigg). 
\end{align}

Since $\alpha,~\beta$, and $\delta$ are the discrete parameters, maximization of the log--likelihood function with respect to these parameters cannot be carried out through ordinary differentiations. However, their parameter space is finite, so all possible combinations can be considered explicitly. For each fixed combination of the discrete parameters 
\[
(\delta,\alpha,\beta)\in
\{-1,1\}\times \mathbf{Z_{m_1}}\times \mathbf{Z_{m_2}},
\]
the log--likelihood function given in \eqref{eq:6.1} becomes a function of the continuous parameters $q,~s,~\rho$. Now we maximize the resulting function with respect to the continuous parameters $q,~s,~\rho$. The corresponding partial derivatives of the log--likelihood are
\begin{align*}
\frac{\partial \ell}{\partial q}
={}& n\frac{\partial \log C_1}{\partial q}
+
\sum_{i=1}^{n}\frac{\zeta_{1_i}q^{\zeta_{1_i}-1}
+
(m_1-\zeta_{1_i})q^{m_1-\zeta_{1_i}-1}}{
q^{\zeta_{1_i}}+q^{m_1-\zeta_{1_i}}}, \\
\frac{\partial \ell}{\partial s}
={}&n\frac{\partial \log C_1}{\partial s}
+
\sum_{i=1}^{n}\frac{\zeta_{2_i}s^{\zeta_{2_i}-1}
+
(m_2-\zeta_{2_i})s^{m_2-\zeta_{2_i}-1}
}{s^{\zeta_{2_i}}+s^{m_2-\zeta_{2_i}}}, \\
\frac{\partial \ell}{\partial \rho}
={}&n\frac{\partial \log C_1}{\partial \rho}
+\sum_{i=1}^{n}\frac{c_i}{1+\rho c_i}, 
\end{align*}
where
$c_i=
\cos\left(
\frac{2\pi\zeta_{1_i}}{m_1}
-\delta\frac{2\pi\zeta_{2_i}}{m_2}
\right).
$

The resulting expressions do not lead to closed-form estimates. Therefore, the log--likelihood is maximized numerically using the \texttt{L-BFGS-B} algorithm, implemented through the built-in R function \texttt{optim}. Thus, $2m_1m_2$ numerical optimizations are performed. The discrete parameter combination and the corresponding continuous parameter estimates yielding the largest maximized log-likelihood are selected as the maximum likelihood estimate. For the selected combination of discrete parameters $(\delta,\alpha,\beta)$, let
\[\boldsymbol{\eta}=(q,s,\rho)^\top\]
denote the vector of continuous parameters. The standard error of the $j$th continuous parameter estimate is
computed as
\[
\mathrm{SE}(\widehat{\eta}_j)
=
\sqrt{
\left[
\widehat{\mathrm{Var}}(\widehat{\boldsymbol{\eta}})
\right]_{jj}
},
\]
where
$
\widehat{\mathrm{Var}}(\widehat{\boldsymbol{\eta}})
=
\mathbf{H}_{\mathrm{NLL}}
(\widehat{\boldsymbol{\eta}})^{-1},
$
and $\mathbf{H}_{\mathrm{NLL}}(\widehat{\boldsymbol{\eta}})$ denotes the
Hessian matrix of the negative log-likelihood with respect to the continuous parameters, evaluated at their maximum likelihood estimates.

The parameters of the BGWG distribution are estimated similarly. In this case,
$\delta\in\{-1,1\}$ is the only discrete parameter, whereas
$q,~s,~\rho,~\alpha$, and $\beta$ are continuous. For each fixed value of
$\delta$, the log--likelihood is maximized numerically with respect to
$(q,s,\rho,\alpha,\beta)$.

\section{Simulation study}\label{sec7}
A Monte Carlo simulation study is conducted to assess the finite-sample performance of the proposed MLEs. We fix $m_1=5$ and $m_2=6$, and consider two distinct parameter configurations under the sample sizes $n=50$, $200$, and $500$. For each setting, bivariate random samples are generated sequentially by first sampling $X_1$ from its marginal distribution given in \eqref{eq:3.1} and then sampling $X_2$ from the conditional distribution of $X_2$ given $X_1$, as specified in \eqref{eq:3.3}. Random sampling from these distributions is carried out using the built-in \texttt{sample()} function in \textsf{R}.

\textbf{Results and discussion:} The results are based on $N=1000$ independent simulations. \cref{tab:6.1} and \cref{tab:6.2} present the results for the BWG and BGWG distributions, respectively. The standard errors reported in these tables are the sample standard deviations of the corresponding parameter estimates obtained over the $N$ simulation runs. Since standard errors are not reported for the discrete parameters, the last column(s) of the tables report the frequencies of their estimated values over the $N$ simulation runs. Specifically, for the discrete parameters $\alpha$ and $\beta$, the $i$th entry of the reported frequency vector corresponds to the number of simulation runs in which the discrete parameter is estimated as $i-1$. For $\delta$, the two entries correspond to the number of simulation runs in which $\delta$ is estimated as $-1$ and $1$, respectively.

\begin{table}[!ht]
\caption{Estimates (with standard errors in parentheses) for the BWG distribution}\vspace{-2mm}
\label{tab:6.1}
\centering
\begin{tabular}{ccccccc}
\toprule
 \multicolumn{1}{c}{Sample size} & \multicolumn{6}{c}{MLE (Standard Error)} \\
 \cmidrule(lr){1-7}
 n & $\hat{q}_{\mathrm{MLE}}$  & $\hat{s}_{\mathrm{MLE}}$  & $\hat{\rho}_{\mathrm{MLE}}$ & $\hat{\alpha}$ & $\hat{\beta}$ & $\hat{\delta}$\\
 \hline
 True value & $q=0.2$ & $s=0.3$ & $\rho=-0.5$ & $\alpha=0$ & $\beta=0$ & $\delta=1$\\
 \hline
 50 & 0.203 & 0.299 & -0.453 & (1000,0,0,0,0) & (998,1,0,0,0,1) & (187,813)\\
    & (0.059) & (0.078) & (0.324) & &\\

 200 & 0.201 & 0.300 & -0.493 & (1000,0,0,0,0) & (1000,0,0,0,0,0) & (19,981)\\
     & (0.027) & (0.036) & (0.132) & & &\\

 500 & 0.200 & 0.300 & -0.495 & (1000,0,0,0,0) & (1000,0,0,0,0,0) & (0,1000)\\
     & (0.017) & (0.022) & (0.077) & & &\\
 \hline
 True value & $q=0.6$ & $s=0.7$ & $\rho=0.8$ & $\alpha=2$ & $\beta=2$ & $\delta=-1$ \\
 \hline
 50 & 0.638 & 0.692 & 0.411 & (33,61,796,65,45) & (22,63,712,62,24,117) & (993,7)\\
    & (0.165) & (0.117) & (0.709) & & &\\

 200 & 0.606 & 0.705 & 0.763 & (1,1,995,2,1) & (1,2,977,1,1,18) & (1000,0)\\
     & (0.066) & (0.069) & (0.246) & & &\\

 500 & 0.600 & 0.704 & 0.801 & (0,0,1000,0,0) & (0,0,1000,0,0,0) & (1000,0)\\
     & (0.037) & (0.043) & (0.047) & & &\\
 \hline
\end{tabular}
\end{table}

\begin{table}[!ht]
\caption{Estimates (with standard errors in parentheses) for the BGWG distribution}\vspace{-2mm}
\label{tab:6.2}
\centering
\begin{tabular}{ccccccc}
\toprule
\multicolumn{1}{c}{Sample size} & \multicolumn{6}{c}{MLE (Standard Error)} \\
\cmidrule(lr){1-7}
 n & $\hat{q}_{\mathrm{MLE}}$  & $\hat{s}_{\mathrm{MLE}}$  & $\hat{\rho}_{\mathrm{MLE}}$ & $\hat{\alpha}$ & $\hat{\beta}$ & $\hat{\delta}$\\
 \hline
 True value & $q=0.2$ & $s=0.3$ & $\rho=-0.5$ & $\alpha=2$ & $\beta=3$ & $\delta=1$ \\
 \hline
 50 & 0.175 & 0.276 & -0.476 & 1.999 & 3.002 & (185,815)\\
    & (0.057) & (0.080) & (0.327) & (0.137) & (0.173) &\\

 200 & 0.189 & 0.287 & -0.498 & 2.004 & 2.994 & (8,992)\\
     & (0.027) & (0.035) & (0.119) & (0.070) & (0.091) &\\
 
  500 & 0.193 & 0.293 & -0.494 & 1.997 & 2.999 & (0,1000)\\
      & (0.018) & (0.022) & (0.074) & (0.047) & (0.057)\\
 \hline
 True value & $q=0.6$ & $s=0.7$ & $\rho=0.8$ & $\alpha=3$ & $\beta=2$ & $\delta=-1$\\
 \hline
 50 & 0.558 & 0.667 & 0.330 & 2.869 & 2.482 & (989,11)\\
    & (0.133) & (0.130) & (0.769) & (0.740) & (1.229) &\\ 
 
 200 & 0.585 & 0.692 & 0.756 & 2.988 & 2.074 & (1000,0) \\
     & (0.070) & (0.074) & (0.289) & (0.249) & (0.494) &\\

 500 & 0.589 & 0.693 & 0.800 & 2.996 & 2.011 & (1000,0) \\
     & (0.040) & (0.044) & (0.087) & (0.118) & (0.192) & \\
 \hline
\end{tabular}
\end{table}

From \cref{tab:6.1} and \cref{tab:6.2}, it is clearly seen that as the sample size increases, the estimates become more accurate. The accuracy in estimation of the discrete parameters $\alpha$ and $\beta$, as presented in \cref{tab:6.1}, is notably higher for lower values of parameters $q$ and $s$ (e.g., $q = 0.2, s = 0.3$) compared to larger values (e.g., $q = 0.6, s = 0.7$). This behavior occurs because smaller values of $q$ and $s$ lead to a stronger concentration of the BWG distribution around the location $\left( \frac{2\pi \alpha}{m_1}, \frac{2\pi \beta}{m_2} \right)$. In fact, when $q=0$ and $s=0$, the distribution degenerates exactly at this location. Conversely, as $q$ and $s$ approach 1, the distribution becomes flatter and nearly uniform. These conclusions follow from the limiting cases given in \cref{tab:2.1}. Consequently, the reduced concentration for larger values of $q$ and $s$ results in greater variability in the parameter estimates.

Furthermore, the precision in estimation of the discrete parameter $\delta$, presented in \cref{tab:6.1} and \cref{tab:6.2}, is highly affected by the parameter $\rho$. Specifically, the accuracy is higher for large absolute values of $\rho$ (e.g., $|\rho| = 0.8$) compared to lower absolute values (e.g., $|\rho| = 0.5$). This behavior occurs because the role of $\delta$ in the distributions is governed by the term $\rho\cos\Big(\frac{2\pi \zeta_1}{m_1} - \delta \frac{2\pi \zeta_2}{m_2}\Big)$, and the influence of this term in both distributions directly depends on $|\rho|$. Consequently, as $|\rho|$ decreases, the distinguishability between $\delta=1$ and $\delta=-1$ diminishes.

\section{Data analysis}\label{sec8}
In this section, we apply the proposed models to study the dependence between wind directions recorded at different stations in India. Data were obtained from three locations: Gwalior in north-central India, Chennai on the southeastern coast, and Mangaluru on the southwestern coast. Observations corresponding to periods of calm (no wind) were excluded from the analysis.

\begin{figure}[!ht]
\centering
\includegraphics[width=0.95\linewidth]{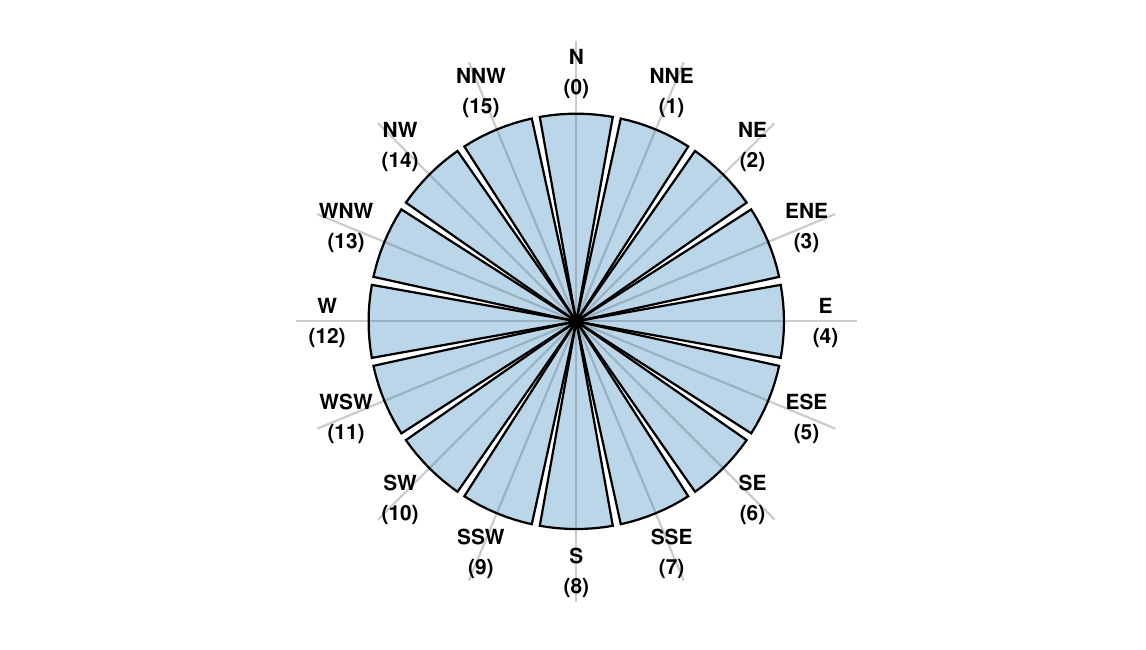}
\caption{Representation of the directions}
\label{fig:7.1}
\end{figure}

The wind directions were recorded in 16 equally spaced compass directions (e.g., N, NNE, NE, $\ldots$, NNW) and were therefore treated as discrete circular observations. These compass directions were indexed by the integers $0,1,\ldots,15$, as illustrated in \cref{fig:7.1}, with the corresponding angular directions represented by $2\pi k/16$, $k=0,1,\ldots,15$. Since the wind directions at each station were recorded in 16 compass directions, we set $m_1=m_2=16$. For each dataset, we fitted two BGWG models, corresponding to $\delta=-1$ and $\delta=1$. The remaining model parameters were estimated by maximum likelihood using the procedure described in \cref{sec6}.

For the interpretation of the fitted BGWG models, $\alpha$ and $\beta$ are the location parameters for the first and second wind-direction variables, respectively, while $q$ and $s$ determine their corresponding concentrations. Since $m_1=m_2=16$, the fitted location parameters $\alpha$ and $\beta$ correspond to the angular locations $2\pi\alpha/16$ and $2\pi\beta/16$, respectively. Smaller values of $q$ and $s$ indicate greater concentration around the corresponding location, whereas values closer to one indicate more dispersed directional observations. The magnitude of $\rho$ describes the strength of dependence between the two wind directions. For $\rho>0$, $\delta=1$ corresponds to rotational dependence and $\delta=-1$ to anti-rotational dependence; when $\rho<0$, the dependence cannot be characterized explicitly as rotational or anti-rotational.

We first analyze the three station pairs---Gwalior--Mangaluru, Chennai--Gwalior, and Chennai--Mangaluru---using the BGWG distribution and interpret the fitted location, concentration, and dependence parameters. We then compare the proposed BWG and BGWG distributions with discretized versions of established continuous bivariate circular models using the Akaike information criterion (AIC).

\subsection{Dataset-I}
\textit{Dataset-I} was constructed from wind direction data collected at two locations: Gwalior Fort in Gwalior and Rao Circle in Mangaluru, during the period from 24 December 2024 to 7 January 2025. \cref{tab:7.2} presents the contingency table for this dataset. 

We fitted the proposed BGWG distribution to \textit{Dataset-I}. Since the model corresponding to $\delta = -1$ had a smaller AIC than the model corresponding to $\delta = 1$, we selected the model with $\delta = -1$ for further analysis. The AIC was computed using the formula \( \mathrm{AIC} = 2\mathcal{P} + 2\mathcal{L} \), where \(\mathcal{P}\) denotes the number of estimated parameters and \(\mathcal{L}\) denotes the negative log-likelihood value. The estimated parameters are provided in \cref{tab:7.1}. \cref{fig:7.2} displays the heatmap of the estimated pmf.

\begin{table}[!ht]
\centering
\small
\caption{Estimated parameters of the BGWG ($\delta = -1$) model for \textit{Dataset-I}}
\label{tab:7.1}
\begin{tabular}{lc}
\toprule
\textbf{Parameter} & \textbf{Estimate (Standard Error)} \\
\midrule
$\hat{\alpha}_{\mathrm{MLE}}$ & 15.138 (0.484) \\
$\hat{\beta}_{\mathrm{MLE}}$  & 15.249 (0.452) \\
$\hat{q}_{\mathrm{MLE}}$      & 0.814 (0.044) \\
$\hat{s}_{\mathrm{MLE}}$      & 0.794 (0.043) \\
$\hat{\rho}_{\mathrm{MLE}}$   & 0.804 (0.123) \\
\bottomrule
\end{tabular}
\end{table}

\begin{table}[!ht]
\centering
 \begin{minipage}[b]{0.65\textwidth}\hspace{-48mm} 
  \centering
  \resizebox{0.8\textwidth}{0.18\textheight}{ 
   \begin{tabular}{cc|cccccccccccccccc}
    \multirow{16}{*}{\rotatebox[origin=c]{90}{\textbf{Rao Circle}}}
    & 15 & 0 & 1 & 0 & 0 & 1 & 1 & 0 & 0 & 0 & 0 & 0 & 0 & 0 & 0 & 1 & 3 \\ 
    & 14 & 2 & 0 & 3 & 0 & 0 & 0 & 0 & 0 & 0 & 0 & 0 & 0 & 0 & 1 & 1 & 0 \\
    & 13 & 1 & 1 & 2 & 2 & 1 & 1 & 0 & 0 & 0 & 0 & 0 & 0 & 0 & 0 & 2 & 3 \\
    & 12 & 0 & 1 & 2 & 2 & 2 & 1 & 0 & 0 & 0 & 0 & 0 & 0 & 1 & 1 & 2 & 0 \\
    & 11 & 1 & 0 & 0 & 0 & 1 & 0 & 0 & 0 & 0 & 0 & 0 & 0 & 0 & 0 & 1 & 0 \\
    & 10 & 1 & 0 & 0 & 0 & 1 & 0 & 0 & 0 & 0 & 0 & 0 & 0 & 0 & 0 & 0 & 0 \\
    & 9  & 0 & 0 & 0 & 0 & 0 & 0 & 0 & 0 & 0 & 0 & 0 & 0 & 0 & 0 & 0 & 0 \\
    & 8  & 0 & 1 & 0 & 0 & 0 & 0 & 0 & 0 & 0 & 0 & 0 & 0 & 0 & 0 & 0 & 0 \\
    & 7  & 0 & 0 & 0 & 0 & 0 & 0 & 0 & 0 & 0 & 0 & 1 & 0 & 0 & 0 & 0 & 0 \\
    & 6  & 0 & 0 & 0 & 0 & 0 & 0 & 1 & 0 & 0 & 0 & 0 & 0 & 0 & 0 & 0 & 0 \\
    & 5  & 0 & 0 & 0 & 1 & 0 & 0 & 0 & 1 & 0 & 0 & 1 & 0 & 0 & 0 & 0 & 0 \\
    & 4  & 0 & 0 & 0 & 0 & 0 & 1 & 3 & 0 & 0 & 0 & 2 & 0 & 0 & 4 & 2 & 0 \\
    & 3  & 0 & 0 & 1 & 0 & 0 & 1 & 0 & 0 & 1 & 1 & 0 & 1 & 1 & 2 & 1 & 0 \\
    & 2  & 1 & 0 & 1 & 0 & 0 & 1 & 0 & 0 & 0 & 0 & 0 & 1 & 0 & 0 & 0 & 1 \\
    & 1  & 0 & 0 & 0 & 1 & 0 & 0 & 0 & 0 & 0 & 0 & 0 & 0 & 0 & 1 & 1 & 1 \\
    & 0  & 3 & 0 & 2 & 1 & 0 & 1 & 0 & 0 & 0 & 1 & 0 & 0 & 0 & 2 & 1 & 2 \\
    \midrule 
    \multicolumn{2}{c|}{} & 0 & 1 & 2 & 3 & 4 & 5 & 6 & 7 & 8 & 9 & 10 & 11 & 12 & 13 & 14 &15\\
    \multicolumn{2}{c|}{} & \multicolumn{16}{c}{\textbf{Gwalior Fort}} \\  
   \end{tabular}
  }
  \par\vspace{5pt}\hspace{-50mm} 
  \parbox{\linewidth}{\caption{Contingency table of \textit{Dataset-I}} \label{tab:7.2}}
\end{minipage}\hspace{-34mm}
\begin{minipage}[b]{0.3\textwidth} 
  \centering
  \includegraphics[width=64mm, height=76mm]{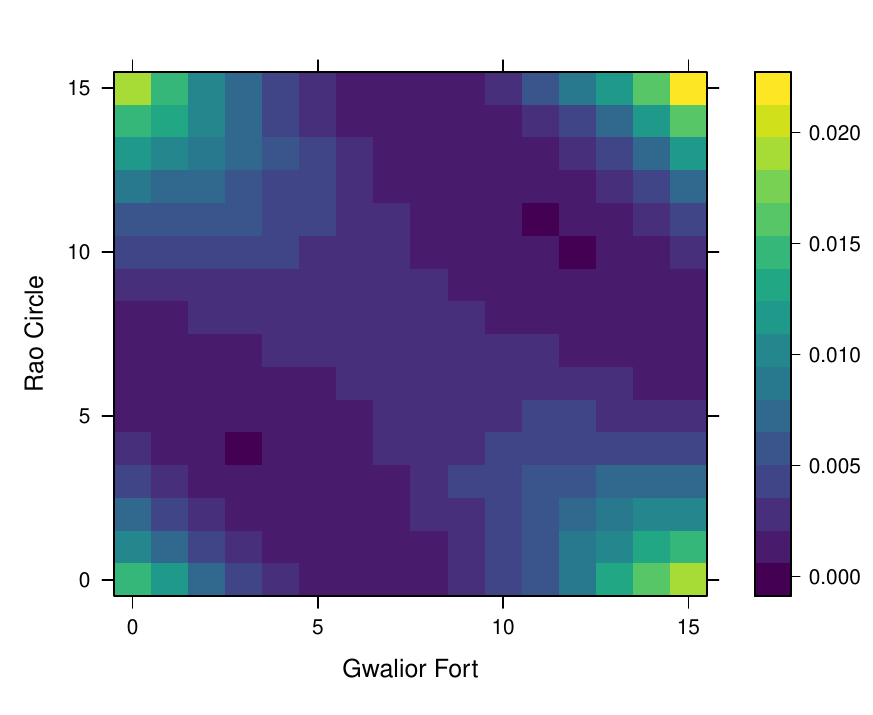} 
\par\vspace{-5mm}
 \captionsetup{justification=centering,format=plain} 
  \captionof{figure}{\hspace{6mm}\makebox[0.4\linewidth]{Heatmap of fitted} BGWG $(\delta=-1)$ pmf}
\vspace{-3mm}
  \label{fig:7.2}
\end{minipage}
\end{table}

\noindent\textbf{Interpretation:}
For Dataset-I, the estimates $\hat{\alpha}_{\mathrm{MLE}}=15.138$ and
$\hat{\beta}_{\mathrm{MLE}}=15.249$ correspond to fitted angular locations
$15.138\pi/8$ and $15.249\pi/8$ for Gwalior Fort and Rao Circle,
respectively. Thus, the fitted locations at both stations lie between NNW
$(15\pi/8)$ and N $(2\pi)$, close to the NNW direction. The estimates
$\hat{q}_{\mathrm{MLE}}=0.814$ and $\hat{s}_{\mathrm{MLE}}=0.794$ indicate
relatively low concentration of the wind directions around the corresponding
fitted locations at Gwalior Fort and Rao Circle. The estimate
$\hat{\rho}_{\mathrm{MLE}}=0.804$ indicates a strong directional association between the wind directions at the two stations. Since $\hat{\rho}_{\mathrm{MLE}}>0$ and the selected model has $\hat{\delta}_{\mathrm{MLE}}=-1$, this association is anti-rotational. Thus, relative to their respective fitted locations, clockwise variation in the wind direction at one station tends to be associated with anticlockwise variation at the other station, and vice versa.

\subsection{Dataset-II}
\textit{Dataset-II} was obtained from wind direction observations recorded at two locations: Marina Beach in Chennai and Gwalior Fort in Gwalior, India, during the period from 22 December 2024 to 7 January 2025. \cref{tab:7.4} provides the contingency table for this dataset.

We fitted the BGWG distribution to \textit{Dataset-II}. Among the two fitted models, the model with $\delta = 1$ yielded the smaller AIC and was therefore selected for further analysis. The estimated parameters are provided in \cref{tab:7.3}. \cref{fig:7.3} shows the heatmap of the estimated pmf.

\begin{table}[!ht]
\centering
\small
\caption{Estimated parameters of the BGWG ($\delta = 1$) model for \textit{Dataset-II}}
\label{tab:7.3}
\begin{tabular}{lc}
\toprule
\textbf{Parameter} & \textbf{Estimate (Standard Error)} \\
\midrule
$\hat{\alpha}_{\mathrm{MLE}}$ & 1.421 (0.069) \\
$\hat{\beta}_{\mathrm{MLE}}$  & 1.159 (0.230) \\
$\hat{q}_{\mathrm{MLE}}$      & 0.218 (0.030) \\
$\hat{s}_{\mathrm{MLE}}$      & 0.549 (0.032) \\
$\hat{\rho}_{\mathrm{MLE}}$   & -0.945 (0.036) \\
\bottomrule
\end{tabular}
\end{table}

\begin{table}[!ht]
\centering
 \begin{minipage}[b]{0.65\textwidth}\hspace{-48mm} 
  \centering
  \resizebox{0.8\textwidth}{0.18\textheight}{ 
   \begin{tabular}{cc|cccccccccccccccc}
    \multirow{16}{*}{\rotatebox[origin=c]{90}{\textbf{Gwalior Fort}}}
    & 15 & 0 & 1 & 8 & 1 & 0 & 0 & 0 & 0 & 0 & 0 & 0 & 0 & 0 & 0 & 0 & 0 \\
    & 14 & 1 & 8 & 3 & 0 & 0 & 0 & 0 & 0 & 0 & 0 & 0 & 0 & 0 & 0 & 0 & 0 \\
    & 13 & 0 & 5 & 6 & 1 & 0 & 0 & 0 & 0 & 0 & 0 & 0 & 0 & 0 & 0 & 0 & 0 \\
    & 12 & 0 & 1 & 1 & 0 & 0 & 0 & 0 & 0 & 0 & 0 & 0 & 0 & 0 & 0 & 0 & 0 \\
    & 11 & 0 & 0 & 2 & 0 & 0 & 0 & 0 & 0 & 0 & 0 & 0 & 0 & 0 & 0 & 0 & 0 \\
    & 10 & 0 & 1 & 3 & 0 & 0 & 0 & 0 & 0 & 0 & 0 & 0 & 0 & 0 & 0 & 0 & 1 \\
    & 9  & 0 & 0 & 2 & 0 & 0 & 0 & 0 & 0 & 0 & 0 & 0 & 0 & 0 & 0 & 0 & 0 \\
    & 8  & 1 & 1 & 0 & 1 & 0 & 0 & 0 & 0 & 0 & 0 & 0 & 0 & 0 & 0 & 0 & 0 \\
    & 7  & 0 & 1 & 0 & 0 & 0 & 0 & 0 & 0 & 0 & 0 & 0 & 0 & 0 & 0 & 0 & 0 \\
    & 6  & 1 & 4 & 0 & 2 & 0 & 0 & 0 & 0 & 0 & 0 & 0 & 0 & 0 & 0 & 0 & 0 \\
    & 5  & 1 & 2 & 4 & 1 & 0 & 0 & 0 & 0 & 0 & 0 & 0 & 0 & 0 & 0 & 0 & 2 \\
    & 4  & 2 & 1 & 2 & 2 & 0 & 0 & 0 & 0 & 0 & 0 & 0 & 0 & 0 & 0 & 0 & 0 \\
    & 3  & 2 & 2 & 1 & 1 & 0 & 0 & 0 & 0 & 0 & 0 & 0 & 0 & 0 & 0 & 0 & 2 \\
    & 2  & 5 & 2 & 0 & 0 & 0 & 0 & 0 & 0 & 0 & 0 & 0 & 0 & 0 & 0 & 0 & 4 \\
    & 1  & 2 & 1 & 2 & 0 & 0 & 0 & 0 & 0 & 0 & 0 & 0 & 0 & 0 & 0 & 0 & 0 \\
    & 0  & 1 & 3 & 5 & 0 & 0 & 0 & 0 & 0 & 0 & 0 & 0 & 0 & 0 & 0 & 0 & 0 \\
    \midrule 
    \multicolumn{2}{c|}{} & 0 & 1 & 2 & 3 & 4 & 5 & 6 & 7 & 8 & 9 & 10 & 11 & 12 & 13 & 14 & 15 \\
    \multicolumn{2}{c|}{} & \multicolumn{16}{c}{\textbf{Marina Beach}} \\
   \end{tabular}
  }
  \par\vspace{5pt}\hspace{-50mm} 
  \parbox{\linewidth}{\caption{Contingency table of \textit{Dataset-II}} \label{tab:7.4}}
\end{minipage}\hspace{-34mm}
\begin{minipage}[b]{0.3\textwidth} 
  \centering
  \includegraphics[width=64mm, height=76mm]{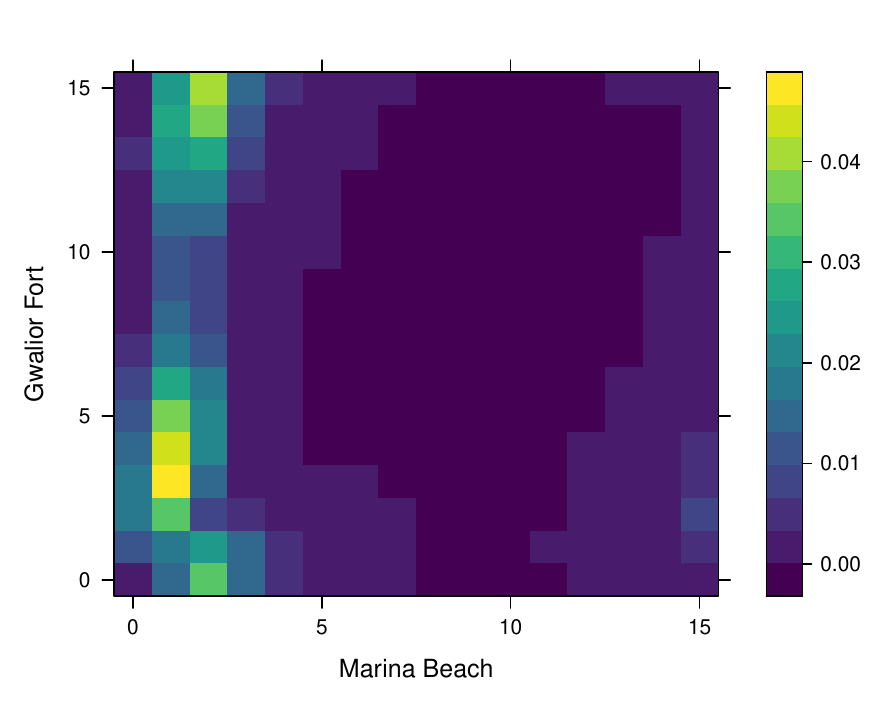} 
\par\vspace{-5mm}
 \captionsetup{justification=centering,format=plain} 
  \captionof{figure}{\hspace{6mm}\makebox[0.4\linewidth]{Heatmap of fitted} BGWG $(\delta=1)$ pmf}
\vspace{-3mm}
  \label{fig:7.3}
\end{minipage}
\end{table}

\noindent\textbf{Interpretation:}
For Dataset-II, the estimates $\hat{\alpha}_{\mathrm{MLE}}=1.421$ and
$\hat{\beta}_{\mathrm{MLE}}=1.159$ correspond to fitted angular locations
$1.421\pi/8$ and $1.159\pi/8$ for Marina Beach and Gwalior Fort,
respectively. Both fitted locations lie between NNE $(\pi/8)$ and NE
$(\pi/4)$, with the fitted location for Gwalior Fort particularly close to
NNE. The estimate $\hat{q}_{\mathrm{MLE}}=0.218$ indicates greater
concentration of the wind directions around the fitted location at Marina
Beach, whereas $\hat{s}_{\mathrm{MLE}}=0.549$ indicates comparatively lower
concentration of the wind directions around the fitted location at Gwalior
Fort. The large magnitude of $\hat{\rho}_{\mathrm{MLE}}=-0.945$ indicates
a strong directional association between the wind directions at the two
stations.

\subsection{Dataset-III}
\textit{Dataset-III} was constructed from wind direction data collected at two locations: Marina Beach in Chennai and Rao Circle in Mangaluru, during the period from 24 December 2024 to 7 January 2025. \cref{tab:7.6} presents the contingency table for this dataset.

We fitted the BGWG distribution to \textit{Dataset-III}. The model with $\delta = 1$ provided the better fit, as indicated by its smaller AIC, and was therefore selected for further analysis. The estimated parameters are provided in \cref{tab:7.5}. \cref{fig:7.4} shows the heatmap of the estimated pmf.

\begin{table}[!ht]
\centering
\small
\caption{Estimated parameters of the BGWG ($\delta = 1$) model for \textit{Dataset-III}}
\label{tab:7.5}
\begin{tabular}{lc}
\toprule
\textbf{Parameter} & \textbf{Estimate (Standard Error)} \\
\midrule
$\hat{\alpha}_{\mathrm{MLE}}$ & 1.523 (0.077) \\
$\hat{\beta}_{\mathrm{MLE}}$  & 8.265 (0.377) \\
$\hat{q}_{\mathrm{MLE}}$      & 0.272 (0.038) \\
$\hat{s}_{\mathrm{MLE}}$      & 0.990 (0.121) \\
$\hat{\rho}_{\mathrm{MLE}}$   & -0.834 (0.101) \\
\bottomrule
\end{tabular}
\end{table}

\begin{table}[!ht]\hspace{-10mm}
\centering
 \begin{minipage}[b]{0.65\textwidth}\hspace{-48mm} 
  \centering
  \resizebox{0.8\textwidth}{0.18\textheight}{ 
   \begin{tabular}{cc|cccccccccccccccc}
    \multirow{16}{*}{\rotatebox[origin=c]{90}{\textbf{Rao Circle}}}
    & 15 & 1 & 2 & 4 & 1 & 0 & 0 & 0 & 0 & 0 & 0 & 0  & 0  & 0  & 0  & 0  & 0  \\
    & 14 & 2 & 2 & 2 & 0 & 0 & 0 & 0 & 0 & 0 & 0 & 0  & 0  & 0  & 0  & 0  & 1  \\
    & 13 & 2 & 2 & 7 & 1 & 0 & 0 & 0 & 0 & 0 & 0 & 0  & 0  & 0  & 0  & 0  & 1  \\
    & 12 & 3 & 1 & 6 & 1 & 0 & 0 & 0 & 0 & 0 & 0 & 0  & 0  & 0  & 0  & 0  & 1  \\
    & 11 & 1 & 1 & 1 & 0 & 0 & 0 & 0 & 0 & 0 & 0 & 0  & 0  & 0  & 0  & 0  & 0  \\
    & 10 & 0 & 0 & 1 & 1 & 0 & 0 & 0 & 0 & 0 & 0 & 0  & 0  & 0  & 0  & 0  & 0  \\
    & 9  & 0 & 0 & 0 & 0 & 0 & 0 & 0 & 0 & 0 & 0 & 0  & 0  & 0  & 0  & 0  & 0  \\
    & 8  & 0 & 0 & 1 & 0 & 0 & 0 & 0 & 0 & 0 & 0 & 0  & 0  & 0  & 0  & 0  & 0  \\
    & 7  & 0 & 0 & 1 & 0 & 0 & 0 & 0 & 0 & 0 & 0 & 0  & 0  & 0  & 0  & 0  & 0  \\
    & 6  & 0 & 0 & 0 & 1 & 0 & 0 & 0 & 0 & 0 & 0 & 0  & 0  & 0  & 0  & 0  & 0  \\
    & 5  & 1 & 1 & 1 & 0 & 0 & 0 & 0 & 0 & 0 & 0 & 0  & 0  & 0  & 0  & 0  & 0  \\
    & 4  & 0 & 7 & 3 & 3 & 0 & 0 & 0 & 0 & 0 & 0 & 0  & 0  & 0  & 0  & 0  & 1  \\
    & 3  & 0 & 6 & 5 & 0 & 0 & 0 & 0 & 0 & 0 & 0 & 0  & 0  & 0  & 0  & 0  & 0  \\
    & 2  & 0 & 1 & 3 & 1 & 0 & 0 & 0 & 0 & 0 & 0 & 0  & 0  & 0  & 0  & 0  & 1  \\
    & 1  & 1 & 2 & 1 & 0 & 0 & 0 & 0 & 0 & 0 & 0 & 0  & 0  & 0  & 0  & 0  & 0  \\
    & 0  & 2 & 3 & 5 & 1 & 0 & 0 & 0 & 0 & 0 & 0 & 0  & 0  & 0  & 0  & 0  & 2  \\
    \midrule 
    \multicolumn{2}{c|}{} & 0 & 1 & 2 & 3 & 4 & 5 & 6 & 7 & 8 & 9 & 10 & 11 & 12 & 13 & 14 & 15 \\
    \multicolumn{2}{c|}{} & \multicolumn{16}{c}{\textbf{Marina Beach}} \\
   \end{tabular}
  }
  \par\vspace{5pt}\hspace{-50mm} 
  \parbox{\linewidth}{\caption{Contingency table of \textit{Dataset-III}} \label{tab:7.6}}
\end{minipage}\hspace{-34mm}
\begin{minipage}[b]{0.3\textwidth} 
  \centering
  \includegraphics[width=64mm, height=76mm]{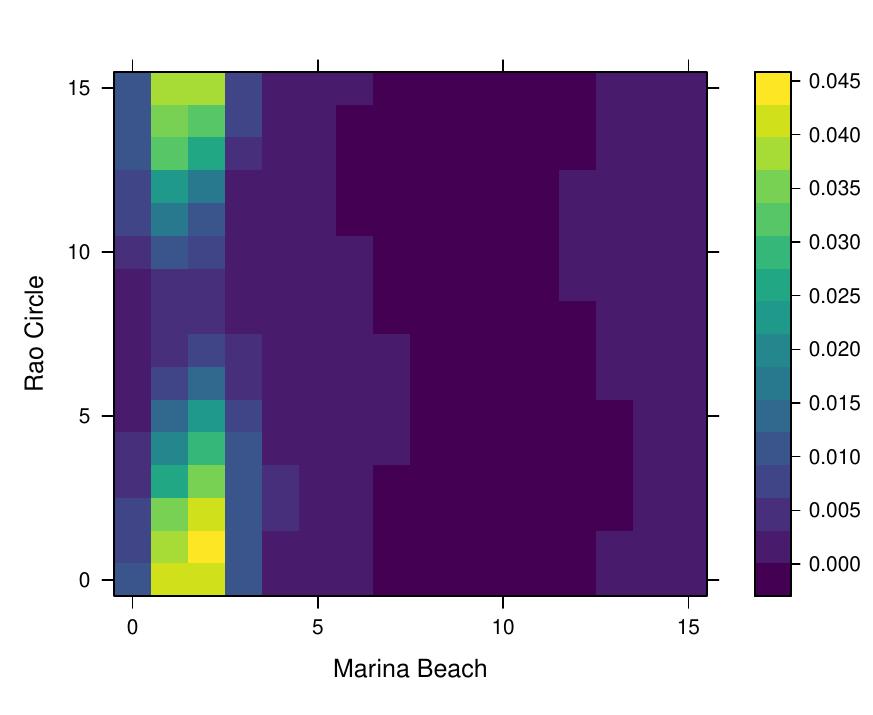} 
\par\vspace{-5mm}
 \captionsetup{justification=centering,format=plain} 
  \captionof{figure}{\hspace{6mm}\makebox[0.4\linewidth]{Heatmap of fitted} BGWG $(\delta=1)$ pmf}
\vspace{-3mm}
  \label{fig:7.4}
\end{minipage}
\end{table}

\noindent\textbf{Interpretation:}
For Dataset-III, the estimates $\hat{\alpha}_{\mathrm{MLE}}=1.523$ and
$\hat{\beta}_{\mathrm{MLE}}=8.265$ correspond to angular locations
$1.523\pi/8$ and $8.265\pi/8$ associated with the fitted location parameters
for Marina Beach and Rao Circle, respectively. The angular location associated
with the fitted location parameter for Marina Beach lies between NNE
$(\pi/8)$ and NE $(\pi/4)$, whereas that for Rao Circle lies close to
S $(\pi)$. The estimate $\hat{q}_{\mathrm{MLE}}=0.272$ indicates greater
concentration associated with the wind directions at Marina Beach. In contrast,
$\hat{s}_{\mathrm{MLE}}=0.990$ is close to one, indicating very low
concentration associated with the wind directions at Rao Circle. The large
magnitude of $\hat{\rho}_{\mathrm{MLE}}=-0.834$ indicates a strong directional
association between the wind directions at Marina Beach and Rao Circle.

\noindent\textbf{Overall interpretation:}
Despite the geographical separation of Gwalior, Chennai, and Mangaluru, the large absolute values of $\hat{\rho}_{\mathrm{MLE}}$ across the three datasets indicate strong directional associations between the wind directions for all three station pairs. This common pattern highlights the relevance of modeling dependence between wind directions recorded at geographically separated stations.

\subsection{Comparison with discretized continuous models}

We compare the performance of the proposed distributions with discretized versions of prominent continuous bivariate distributions: the wrapped Cauchy distribution \citep{kato2015} and the von Mises Sine and Cosine distributions \citep{mardia2007}. For these continuous models, a cell probability discretization was employed, that is, the probability assigned to each point on the discrete torus $\mathbf{D_{16}}\times \mathbf{D_{16}}$ was obtained by integrating the corresponding continuous density over the rectangular cell centered at that point, where each cell has width $\frac{2 \pi}{16}$ in both circular dimensions. The resulting cell probabilities were then used to construct the likelihood for maximum likelihood estimation, and the results are presented in \Cref{tab:7.7}. 

The last row of \cref{tab:7.7} reports the best-fitting BWG model, denoted by  BWG$^{*}$, for each dataset. To obtain these models, BWG models corresponding to all possible combinations of the discrete parameters $(\delta,\alpha,\beta)$ were fitted using the estimation procedure described in \cref{sec6}. For each dataset, the BWG model with the smallest AIC value was selected. The resulting BWG$^{*}$ models correspond to the discrete parameter combinations $(\delta,\alpha,\beta) = (-1,15,15),~ (1,1,1)$ and $(1,2,9)$ for \textit{Dataset-I}, \textit{Dataset-II}, and \textit{Dataset-III}, respectively.

\begin{table}[ht]
\centering
\caption{AIC values for the BWG$^{*}$ and BGWG distributions and the discretized wrapped Cauchy (D-wC), von Mises Sine (D-vM-Sine), and von Mises Cosine (D-vM-Cosine) models fitted to the three wind direction datasets.}
\label{tab:7.7}
\begin{tabular}{lccc}
\hline
 & \multicolumn{1}{c}{Dataset-I} &
   \multicolumn{1}{c}{Dataset-II} &
   \multicolumn{1}{c}{Dataset-III} \\
\cmidrule(lr){2-2}\cmidrule(lr){3-3}\cmidrule(lr){4-4}
Model &
AIC &
 AIC &
AIC \\
\hline
BGWG $(\delta=-1$) & 975.573 & 899.907 & 826.778\\
BGWG $(\delta=1$) & 1000.089 & 890.961 & \cellcolor{gray!30}822.411\\
D-wC  & 978.609 & 921.386 & 851.526\\
D-vM-Sine & 976.044 & \cellcolor{gray!30}889.110 & 822.844\\
D-vM-Cosine & 992.434 & 889.920 & 822.986\\
BWG$^{*}$  & \cellcolor{gray!30}972.786 & 913.672 & 845.388 \\
\hline
\end{tabular}
\end{table}

\Cref{tab:7.7} summarizes the AIC values for the proposed BWG and BGWG distributions together with those of the discretized wrapped Cauchy and von Mises Sine and Cosine models. For \textit{Dataset-I}, the BWG model with $(-1,15,15)$ provides the best fit, attaining the smallest AIC value among all the fitted models, and the BGWG model with $\delta=-1$ also has a smaller AIC than the discretized models. For \textit{Dataset-II}, the discretized von Mises Sine (D-vM-Sine) model yields the lowest AIC value, indicating the best fit for this dataset. For \textit{Dataset-III}, the BGWG model with $\delta=1$ provides the best fit, as indicated by the smallest AIC among the fitted models.

Overall, the proposed models provide a direct framework for analyzing bivariate discrete circular data and yield fits that are competitive with those of the discretized continuous models. An additional advantage is that their likelihoods can be evaluated directly on the observed discrete support, without the numerical integration required to obtain the corresponding probabilities under the discretized continuous models.

Goodness-of-fit tests were conducted for all datasets using Pearson’s chi-square statistic, and the corresponding results are provided in the supplementary material. 

\section{Conclusion}\label{sec9}
This study addresses the challenges associated with modeling bivariate discrete circular random variables. 
We proposed new models defined on the discrete torus and examined how their parameters influence the underlying dependence structure. 
The performance of these models was evaluated using three bivariate wind direction datasets and compared with the discretized versions of well-known existing models. 
The proposed framework can also be applied to analyze how wave directions at one location correlate with those at other locations in the ocean, with potential applications in areas such as animal movement, traffic flow, and seismic activity. 

Using the proposed bivariate models, we analyzed the joint wind direction observations from pairs of meteorological stations. Our findings demonstrate a strong relationship between the wind directions at these stations, even though they are geographically distant, possibly reflecting the effect of the geography of the Indian subcontinent. In the future, multivariate extensions could be constructed using copula-based frameworks to link the proposed discrete circular marginals with continuous linear variables such as wind speed and temperature. Furthermore, time-series analysis can be a generalization of the proposed framework, potentially utilizing Hidden Markov Models to capture temporal dependence.

\section*{Declarations}
\textbf{Funding} \\
This project is partially funded by the Government of India as part of the Start-up Research Grant. The funding is provided through the Science and Engineering Research Board, Department of Science and Technology, India to Dr. Jayant Jha (SRG/2022/000151). Brajesh Kumar Dhakad is supported by the National Board for Higher Mathematics (NBHM) (0203/4/2022/R\&D-II/ 3262). The funders had no role in the methodological development, data analysis, decision to publish, or preparation of the manuscript.

\noindent
\textbf{Competing Interests} \\
The authors have no relevant financial or non-financial interests to disclose.

\noindent
\textbf{Data Availability} \\
The data supporting the findings of this study are available from
\url{https://drive.google.com/file/d/1z4tW084oqywaz-P59672Vm0CVOq38KZD/view?usp=sharing}.

\noindent
\textbf{Code Availability} \\
The R codes used for the data analysis are provided in the supplementary material.


\bibliography{sn-bibliography}

\begin{appendices}

\section{Proof of \texorpdfstring{\cref{theo.2.1}}{Theorem 2.1}}
\label{app1}
From equation~\eqref{eq:2.3}, the reciprocal of $C_1$ is obtained as follows:
 \vspace{-2mm}
\begin{equation*}
\scalebox{0.85}{$
\begin{aligned}   
 \frac{1}{C_1} & = \sum_{k = 0}^{{m_1}-1}\sum_{l=0}^{{m_2}-1}\big(q^{\zeta_1} + q^{{m_1}-\zeta_1}\big)\big(s^{\zeta_2} + s^{{m_2}-\zeta_2}\big)\bigg(1 + \rho \cos\Big(\frac{2\pi \zeta_1}{m_1} - \delta \frac{2\pi \zeta_2}{m_2}\Big)\bigg)\\
&=\big(1-q^{m_1}\big)\big(1-s^{m_2}\big)\big(1+q\big)\big(1+s\big)\Bigg(\frac{1}{\big(1-s \big)\big(1-q\big)} + \rho\frac{\big(1-q \big)\big(1-s \big)}{\big(q^2 -2q \cos\frac{2\pi}{m_1} +1\big)\big(s^2 -2s \cos\frac{2\pi}{m_2} + 1\big)}\Bigg).  
\end{aligned}
$}
\end{equation*}
The closed-form expression of the double sum is obtained by first expressing \( \cos(\cdot) \) in terms of exponential functions using Euler’s formula, and then applying the finite geometric series. 

The function \eqref{eq:2.5} is non-negative and its values sum to one; hence it serves as the pmf of $(X_1, X_2)$.

\section{Proof of \texorpdfstring{\cref{coro.3.1}}{Corollary 3.1}}
\label{app2}
\vspace{-3mm}
Setting $\rho=0$ in Theorem~\ref{theo.2.1}, the joint pmf in \eqref{eq:2.5} reduces to
\begin{equation}
\label{eq:a1}
\scalebox{0.85}{$
\begin{aligned}
 P\!\left(\frac{2\pi k}{m_1},\frac{2\pi l}{m_2}\right) & = 
 \left(\frac{\left(1-q\right)\left(1-s\right)}{\left(1-q^{m_1}\right)\left(1-s^{m_2}\right)\left(1+q\right)\left(1+s\right)}\right)\left(q^{\zeta_1} + q^{m_1-\zeta_1}\right)\left(s^{\zeta_2} + s^{m_2-\zeta_2}\right).
\end{aligned}
$}
\end{equation}
The marginal pmf of $X_1$ is obtained by summing the pmf in \eqref{eq:a1} over the values {\small $\frac{2\pi l}{m_2}$}, where $l \in \mathbf{Z_{m_2}}$, and is given by:

\begin{align}\label{eq:a2}
P\bigg( \frac{2\pi k}{m_1}\bigg) & = \bigg(\frac{1-q}{(1-q^{m_1})(1+q)}\bigg)\Big(q^{\zeta_1} +q^{m_1-\zeta_1}\Big).   
\end{align}
Similarly, the marginal pmf of $X_2$ can be derived, and it is given by:
\begin{equation}
\label{eq:a3}
\resizebox{0.55\textwidth}{!}{$
\begin{aligned}
P\!\left( \frac{2\pi l}{m_2}\right)& = \left(\frac{1-s}{(1-s^{m_2})(1+s)}\right)\left(s^{\zeta_2} +s^{m_2-\zeta_2}\right).   
\end{aligned}
$}
\end{equation}
Hence, the joint pmf in \eqref{eq:a1} factorizes into the product of the marginals given in \eqref{eq:a2} and \eqref{eq:a3}.
Hence,
\[
P\!\left(\frac{2\pi k}{m_1},\frac{2\pi l}{m_2}\right)
=
P\!\left(\frac{2\pi k}{m_1}\right)
P\!\left(\frac{2\pi l}{m_2}\right),
\]
which proves that $X_1$ and $X_2$ are independent when $\rho=0$.

Conversely, suppose that $X_1$ and $X_2$ are independent. Then the joint pmf in
\eqref{eq:2.5} must factorize into the product of a function of $\zeta_1$ and a
function of $\zeta_2$. However, for $\rho\neq0$, the interaction term
\[
1+\rho\cos\left(\frac{2\pi\zeta_1}{m_1}
-\delta\frac{2\pi\zeta_2}{m_2}\right)
\]
cannot be expressed as the product of a function of $\zeta_1$ and a function of
$\zeta_2$. Indeed,
\[
\cos(a-b)=\cos a\cos b+\sin a\sin b,
\]
which depends jointly on both variables $a$ and $b$ and is therefore not separable. Consequently, for $q,s\in(0,1)$, the joint pmf cannot factorize into the product of its marginals unless $\rho=0$. Hence, $X_1$ and $X_2$ are independent only if $\rho=0$.

\section{Proof of \texorpdfstring{\cref{theo.3.2}}{Theorem 3.2}}
\label{app3}
Let {\small $f_1(k) = (q^{\zeta_1} + q^{m_1-\zeta_1}),~ f_2(l) = (s^{\zeta_2} + s^{m_2-\zeta_2}),\,f_3\big(\frac{2\pi k}{m_1},\frac{2\pi l}{m_2}\big) = f_1(k)f_2(l)
 \big(1 \\+ \rho \cos\big(\frac{2\pi \zeta_1}{m_1} - \delta \frac{2\pi \zeta_2}{m_2}\big)\big)$}, and
 $f(x) = a^x + a^{m-x}$, $x\in[0, m)$, where, $a \in (0,1)$, $m\in \mathbf{Z_+}$ are constants. Then, 
\begin{align*}
  f(x)-f(0) = (a^x-1)(1-a^{m-x}) < 0,\quad \forall x \in (0, m).  
\end{align*}

This implies that {\small $f(x) < f(0)$}. Furthermore, it follows that

\begin{equation} 
\label{eq:a5}
\scalebox{0.85}{$
\begin{aligned}
f_1(k) < f_1(\alpha),\quad \forall k\in \mathbf{Z_{m_1}}/\{\alpha\}\quad \text{and}\quad f_2(l) < f_2(\beta),\quad \forall l \in \mathbf{Z_{m_2}}/\{\beta\}. 
\end{aligned} 
$}
\end{equation}

From equation \eqref{eq:a5}, if $\rho\geq0$. Then {\small $ \forall (k,l) \in \mathbf{Z_{m_1}} \times\mathbf{Z_{m_2}}/\{(\alpha,\beta)\} $},

\begin{equation}
\label{eq:a6}
\resizebox{0.85\textwidth}{!}{$
\begin{aligned}
 f_3\bigg(\frac{2\pi k}{m_1},\frac{2\pi l}{m_2}\bigg)& < f_1(\alpha)f_2(\beta)\bigg(1 + \rho \cos\bigg(\frac{2\pi \zeta_1}{m_1} - \delta  \frac{2\pi \zeta_2}{m_2}\bigg)\bigg) \leq f_3\bigg(\frac{2\pi \alpha}{m_1},\frac{2\pi\beta}{m_2}\bigg),
\end{aligned}
$}
\end{equation}
which implies that $f_3\Big(\frac{2\pi \alpha}{m_1},\frac{2\pi\beta}{m_2}\Big) $ is the highest value of $f_3(\cdot)$. Therefore, from equation  \eqref{eq:a6} and the pmf given in \eqref{eq:2.5}, the BWG distribution has a dominant mode at the unique location {\small $ \big(\frac{2\pi \alpha}{m_1},\frac{2\pi\beta}{m_2}\big) $}.

\section{Proof of \texorpdfstring{\cref{theo.3.3}}{Theorem 3.3}}
\label{app4}

If $q=0$, then the MWSG distribution degenerates at
$\frac{2\pi\alpha}{m_1}$. For $0<q<1$, we take $\alpha=0$ without
loss of generality.

For $k<\frac{m_1-1}{2}$, we have
\[
\cos\frac{2\pi k}{m_1}
\geq
\cos\frac{2\pi(k+1)}{m_1},
\]
since cosine decreases on $[0,\pi]$. Moreover,
\begin{align*}
&\left(q^k+q^{m_1-k}\right)
-\left(q^{k+1}+q^{m_1-(k+1)}\right)\\
&\qquad
=(1-q)\left(q^k-q^{m_1-k-1}\right)>0,
\end{align*}
because $k<\frac{m_1-1}{2}$ implies
$k<m_1-k-1$ and $0<q<1$. Hence,
\[
q^k+q^{m_1-k}
>
q^{k+1}+q^{m_1-(k+1)}.
\]

Now, let
\[
B_k=
s^2-2s\cos\left(\frac{2\pi}{m_2}\right)+1
+\rho(1-s)^2\cos\left(\frac{2\pi k}{m_1}\right).
\]
Since $\rho\geq0$, the above cosine inequality gives
\[
B_k\geq B_{k+1}.
\]
Also,
\begin{align*}
B_k
&\geq
s^2-2s\cos\left(\frac{2\pi}{m_2}\right)+1
-\rho(1-s)^2\\
&=
(1-\rho)(1-s)^2
+2s\left(1-\cos\left(\frac{2\pi}{m_2}\right)\right)
\geq0.
\end{align*}
Therefore, from the pmf given in \eqref{eq:3.1},
\begin{align}
\label{eq:a9}
P\left(\tfrac{2\pi k}{m_1}\right)
\geq
P\left(\tfrac{2\pi(k+1)}{m_1}\right),
\qquad
k<\tfrac{m_1-1}{2}.
\end{align}

Likewise, when $k\geq\frac{m_1-1}{2}$, the reverse inequalities
hold, namely,
\[
q^k+q^{m_1-k}
\leq
q^{k+1}+q^{m_1-(k+1)}
\]
and
\[
B_k\leq B_{k+1}.
\]
Since both factors are non-negative, it follows that
\begin{align}
\label{eq:a10}
P\left(\tfrac{2\pi k}{m_1}\right)
\leq
P\left(\tfrac{2\pi(k+1)}{m_1}\right),
\qquad
k\geq\tfrac{m_1-1}{2}.
\end{align}

Thus, for $0<q<1$, the pmf decreases as the circular distance from $k=0$ increases, and hence the MWSG distribution is unimodal with mode at $\frac{2\pi\alpha}{m_1}$. For $q=1$, the limiting distribution is uniform when $s=1$ or $\rho=0$; otherwise, for $s<1$ and $\rho>0$, it is uniquely maximized at $\zeta_1=0$. Therefore, excluding these uniform limiting cases, the MWSG distribution is unimodal with mode at $\frac{2\pi\alpha}{m_1}$.

\section{Proof of \texorpdfstring{\cref{coro.4.1}}{Corollary 4.1}}
\label{app5}
The proof is analogous to that of \cref{coro.3.1}.

\section{Proof of \texorpdfstring{\cref{theo.5.1}}{Theorem 5.1}}
\label{app6}
The proof is carried out by examining the two cases $\delta = 1$ and $\delta = -1$. We first consider the case $\delta = 1$. We set $\alpha = 0$ and $\beta = 0$ without loss of generality to simplify the computation of $\rho_1^2$. Under this assumption, the covariances between $\cos X_1$ and $\sin X_2$, $\sin X_1$ and $\cos X_2$, and $\cos X_2$ and $\sin X_2$ are all zero. Consequently, $\rho_{1cs}$, $\rho_{1sc}$, $\rho'_1$ and $\rho'_2$ also vanish, and $\rho^2_1$ reduces to the following simplified form:

\begin{align}\label{eq:a11}
\rho^2_1 = \rho^2_{1cc} + \rho^2_{1ss}.  
\end{align}
We next compute $\rho^2_{1cc}$ and $\rho^2_{1ss}$ using the squared circular correlation measures \[ 
\rho^2_{1cc}
 = \frac{\operatorname{cov}(\cos X_1,\cos X_2)^2}{\operatorname{var}(\cos X_1)\operatorname{var}(\cos X_2)} \quad\text{and}\quad \rho^2_{1ss}
 = \frac{\operatorname{cov}(\sin X_1,\sin X_2)^2}{\operatorname{var}(\sin X_1)\operatorname{var}(\sin X_2)}, \] where $\operatorname{var}(\cdot)$ and $\operatorname{cov}(\cdot)$ denote the variance and covariance, respectively. The required trigonometric moments are given below:

\begin{equation*}
\scalebox{0.85}{$
\begin{aligned}
\operatorname{var}(\sin X_1) = &\frac{1}{2}(1-q)A_1H_1\Bigg(B_1\bigg(\frac{1}{1-q} - \frac{1-q}{A_2}\bigg) + (1-q)(1-s)^2\bigg(\frac{1}{A_1} - \frac{1}{2}\bigg(\frac{1}{A_3} +\frac{1}{A_1}\bigg)\bigg)\rho\Bigg),\\    
\operatorname{var}(\sin X_2) = &\frac{1}{2}(1-s)B_1H\Bigg(A_1\bigg(\frac{1}{1-s} - \frac{1-s}{B_2}\bigg) + (1-q)^2(1-s)\bigg(\frac{1}{B_1} - \frac{1}{2}\bigg(\frac{1}{B_3} + \frac{1}{B_1}\bigg)\bigg)\rho\Bigg),\\
\operatorname{var}(\cos X_1) = &(1-q)A_1H^2\Bigg(\frac{A_1B^2_1}{2}\bigg(\frac{1}{1-q} + \frac{1-q}{A_2}\bigg) - \frac{B_1^2(1-q)^3}{A_1} + \bigg(\frac{A_1B_1(1-q)(1-s)^2 }{4}\bigg(\frac{1}{A_3} + \frac{3}{A_1}\bigg) - \frac{B_1(1-q)^2}{2} \\ (1-s)^2&\bigg(\frac{1}{1-q} + \frac{1-q}{A_2}\bigg)\bigg)\rho - \frac{1}{4}\bigg(A_1(1-q)(1-s)^4 \bigg(\frac{1}{1-q} + \frac{1-q}{A_2}\bigg)^2 - (1-q)^3(1-s)^4\bigg(\frac{1}{A_3} + \frac{3}{A_1}\bigg)\bigg)\rho^2 \Bigg),\\
\operatorname{var}(\cos X_2) = &(1-s)B_1H^2\Bigg(\frac{A_1^2B_1}{2}\bigg(\frac{1}{1-s} + \frac{1-s}{B_2}\bigg) - \frac{A_1^2(1-s)^3}{B_1} + \bigg(\frac{A_1B_1}{4}(1-q)^2(1-s) \bigg(\frac{1}{B_3}+\frac{3}{B_1}\bigg) -  \frac{A_1(1-s)^2}{2} \\ (1-q)^2&\bigg(\frac{1}{1-s} + \frac{1-s}{B_2}\bigg)\bigg)\rho - \frac{1}{4}\bigg(B_1(1-q)^4(1-s) \bigg(\frac{1}{1-s} + \frac{1-s}{B_2} \bigg)^2 - (1-q)^4(1-s)^3\bigg(\frac{1}{B_3} + \frac{3}{B_1}\bigg)\bigg)\rho^2 \Bigg), \\ 
\end{aligned}
$}
\end{equation*}
\vspace{-3mm}
\begin{equation*}
\scalebox{0.85}{$
\begin{aligned}
\operatorname{cov}(\sin X_1,\sin X_2) = &\frac{1}{4}(1-q)(1-s)A_1B_1H\Bigg(\frac{1-s}{B_2} - \frac{1}{1-s}\Bigg)\Bigg(\frac{1-q}{A_2} - \frac{1}{1-q}\Bigg)\rho, \\ 
\operatorname{cov}(\cos X_1,\cos X_2) = &(1-q)(1-s)H^2\Bigg( 
\frac{A_1^2}{2}\bigg(\frac{1}{1-q} + \frac{1-q}{A_2}\bigg) -(1-q)^3\bigg)\Bigg) \Bigg(\frac{B_1^2}{2} \bigg(\frac{1}{1-s} + \frac{1-s}{B_2}\bigg) - (1-s)^3\bigg)\Bigg)\rho,   
\end{aligned}
$}
\end{equation*}
where
\begin{align*}
\frac{1}{H}& = A_1B_1+\rho(1-q)^2(1-s)^2,~
A_i& = 1+q^2 -2q\cos\frac{2i\pi}{m_1},~
B_i& = 1 + s^2 - 2s\cos\frac{2i\pi}{m_2},
\end{align*}
for ${i\in \{1,2,3\}}$. Therefore, these expressions are used to calculate $\rho^2_1$ for BWG distribution. Let $f(\rho) = \rho^2_{1ss}$ and $g(\rho) = \rho^2_{1cc}$. Then,
\vspace{-2mm}
\begin{equation}\label{eq:a12} 
\resizebox{0.4\textwidth}{!}{$
\begin{aligned}
 f(\rho) = \frac{\alpha_3\rho^2}{(B_1\alpha_1 +\alpha_4\rho)(A_1\alpha_2 + \alpha_5\rho)},  
 \end{aligned}
 $}
\end{equation}
where
\begin{align*}
\alpha_1& = \Big(\frac{1}{1-q} - \frac{1-q}{A_2}\Big),~\alpha_2 = \Big(\frac{1}{1-s} - \frac{1-s}{B_2}\Big),~\alpha_3 = \frac{1}{4}(1-q)(1-s)A_1B_1\alpha^2_1\alpha^2_2,\\
\alpha_4& = (1-q)(1-s)^2\Big(\frac{1}{A_1} - \frac{1}{2}\Big(\frac{1}{A_3} + \frac{1}{A_1}\Big)\Big),~\alpha_5 = (1-q)^2(1-s)\Big(\frac{1}{B_1} - \frac{1}{2}\Big(\frac{1}{B_3} + \frac{1}{B_1}\Big)\Big), 
\end{align*}
and
\begin{equation}\label{eq:a13}
\begin{aligned}
g(\rho) = \frac{(1-q)(1-s)A_1B_1\beta^2_1\beta^2_2\rho^2}{(B_1^2\beta_1 + \beta_3\rho - \beta_5\rho^2)(A_1^2\beta_2 + \beta_4\rho - \beta_6\rho^2)},
\end{aligned}
\end{equation}
where 
\begin{align*}
\beta_1& = \frac{A_1}{2}\Big(\frac{1}{1-q} + \frac{1-q}{A_2}\Big) - \frac{(1-q)^3}{A_1},\\ \beta_2& = \frac{B_1}{2}\Big(\frac{1}{1-s} + \frac{1-s}{B_2}\Big) - \frac{(1-s)^3}{B_1},\\ \beta_3& = \frac{B_1}{2}(1-q)(1-s)^2\Big(\frac{A_1}{2}\Big(\frac{1}{A_3} + \frac{3}{A_1}\Big) - (1-q)\Big(\frac{1}{1-q} + \frac{1-q}{A_2}\Big)\Big),\\
\beta_4& = \frac{A_1}{2}(1-q)^2(1-s)\Big(\frac{B_1}{2}\Big(\frac{1}{B_3} + \frac{3}{B_1}\Big) - (1-s)\Big(\frac{1}{1-s} + \frac{1-s}{B_2}\Big)\Big),\\
\beta_5& = \frac{1}{4}(1-q)(1-s)^4\Big(A_1\Big(\frac{1}{1-q} + \frac{1-q}{A_2}\Big)^2 - (1-q)^2\Big(\frac{1}{A_3} + \frac{3}{A_1}\Big)\Big),\\
\beta_6& = \frac{1}{4}(1-q)^4(1-s)\Big(B_1\Big(\frac{1}{1-s} + \frac{1-s}{B_2}\Big)^2 - (1-s)^2\Big(\frac{1}{B_3} + \frac{3}{B_1}\Big)\Big).
\end{align*} 
Here, $f(\rho)$ and $g(\rho)$ are the ratios of two polynomials in $\rho$. The non-negativity property of variance implies the expressions $B_1\alpha_1 +\alpha_4\rho,~A_1\alpha_2 + \alpha_5\rho,~B_1^2\beta_1 + \beta_3\rho - \beta_5\rho^2$, and $A_1^2\beta_2 + \beta_4\rho - \beta_6\rho^2$ are non-negative. Additionally, the cases where these factors equal zero are independent of the discussion here. Thus, the denominators of $f(\rho)$ and $g(\rho)$ are positive, and therefore, $f(\rho)$ and $g(\rho)$ are differentiable functions with respect to $\rho$. The derivative of $f(\rho)$ with respect to $\rho$, denoted $f'(\rho)$, is given by
\begin{equation}
\label{eq:a14}
\resizebox{0.6\textwidth}{!}{$
\begin{aligned}
f'(\rho) = \frac{\alpha_3\rho\big(B_1\alpha_1(A_1\alpha_2 + \alpha_5\rho) + A_1\alpha_2(B_1\alpha_1 + \alpha_4\rho))}{\big((B_1\alpha_1 +\alpha_4\rho)(A_1\alpha_2 + \alpha_5\rho))^2}.
\end{aligned}
$}
\end{equation}

 Since $A_1$ and $B_1$ are non-negative, it follows that $\alpha_1$ and $\alpha_2$ are also non-negative. Therefore, for $\rho \geq 0$,  $ f'(\rho)$ is non-negative, and hence $f(\rho)$ is monotonically increasing in $\rho$.
 The derivative of $g(\rho)$ with respect to $\rho$, denoted $g'(\rho)$, is given by
 \begin{equation}
 \resizebox{0.9\textwidth}{!}{$
\begin{aligned}\label{eq:a15}
 g'(\rho)& = \Big\{(1-q)(1-s)A_1B_1\beta^2_1\beta^2_2\rho\big[(B_1^2\beta_1 + \beta_5\rho^2)(A_1^2\beta_2 + \beta_4\rho - \beta_6\rho^2) + (B_1^2\beta_1 + \beta_3\rho \\
 &\quad - \beta_5\rho^2)(A_1^2\beta_2 + \beta_6\rho^2)\Big]\Big\}\Big/\Big\{\big[(B_1^2\beta_1 + \beta_3\rho - \beta_5\rho^2)(A_1^2\beta_2 + \beta_4\rho - \beta_6\rho^2)\big]^2\Big\}.   
\end{aligned}
$}
\end{equation}
If $B_1^2\beta_1 + \beta_5\rho^2$ and $A_1^2\beta_2 + \beta_6\rho^2$ are non-negative, then $g'(\rho)$ is non-negative, and hence $g(\rho)$ is monotonically increasing in $\rho$. Therefore, it suffices to show that these two expressions are non-negative. 
Now,
\begin{equation}\label{eq:a16}
\scalebox{0.85}{$
\begin{aligned}
B_1^2\beta_1 + \beta_5\rho^2 \geq &(1-s)^4\rho^2 \Bigg(\frac{A_1}{2}\Big(\frac{1}{1-q} + \frac{1-q}{A_2}\Big) - \frac{(1-q)^3}{A_1} + \frac{1}{4}(1-q)\bigg(A\Big(\frac{1}{1-q} + \frac{1-q}{A_2}\Big)^2 -\\
& (1-q)^2\Big(\frac{1}{A_3} + \frac{3}{A}\Big)\bigg)\Bigg) \nonumber = (1-s)^4\rho^2q(1-x)^3f(x,q),
\end{aligned}
$}
\end{equation}
where 
$
f(x,q)
=
1+Aq+Bq^2+Cq^3+Dq^4+Cq^5+Bq^6+Aq^7+q^8,
$
with
\begin{align*}
A&=2(4+13x+8x^2),\\
B&=4(1+10x+17x^2-8x^3-16x^4),\\
C&=2(4+15x+8x^2-52x^3-144x^4-96x^5),\\
D&=2(3-52x^2-56x^3+184x^4+288x^5+96x^6),
\end{align*}
and $x=\cos\frac{2\pi}{m_1}$.
Since $(1-s)^4\rho^2q(1-x)^3\geq0$, it remains to show that
\[
f(x,q)\geq0,
\qquad x\in[-1,1],~ q\in(0,1).
\]
Let $u=q+\frac{1}{q}.$ Then $q^2+q^{-2}=u^2-2,~
q^3+q^{-3}=u^3-3u$, and $q^4+q^{-4}=u^4-4u^2+2.$ Further define 
\[t=u-2
= q+\frac{1}{q}-2 = \frac{(1-q)^2}{q}.\] Since $q\in(0,1)$, we have $t\geq0$. Substituting $u=t+2$
and simplifying yields

\begin{align}\label{eq:a17}
\frac{f(x,q)}{q^4}
={}&t^4 + 2(8x^2+13x+8)t^3 - 4(16x^4+8x^3-41x^2-49x-18)t^2 \nonumber\\
&+8(1-x^2)(24x^3+68x^2+53x+14)t + 48(1-x^2)^2(2x+1)^2.
\end{align}
We now show that every coefficient in \eqref{eq:a17} is non-negative.

First,
\[
(8x^2+13x+8) > 0, \quad x\in[-1,1],
\]
because its leading coefficient is positive, and its discriminant $13^2-4(8)(8)=-87$ is a negative number. 

Next, let
\[
P_2(x)=16x^4+8x^3-41x^2-49x-18.
\]
We have
$P_2(x)+2 = (x+1)R(x),$ where 
$R(x)=16x^3-8x^2-33x-16.$
 Derivative of $R(x)$ is
\[
R'(x)=48x^2-16x-33,
\]
whose roots are
\[
x=\frac{1}{6}\pm\frac{\sqrt{103}}{12}.
\]
Only
\[
x_0=\frac{1}{6}-\frac{\sqrt{103}}{12}
\]
belongs to $[-1,1]$. Moreover, $R(-1)=-7,~ R(1)=-41,$
and
\[
R(x_0)
=
\frac{-1169+103\sqrt{103}}{54}<0.
\]
Hence $R(x)<0,~ x\in[-1,1].$
Since $x+1\geq0$ on $[-1,1]$, it follows that
$ P_2(x)+2\leq0,$ and therefore $P_2(x)\leq-2<0.$
Consequently,
\[
-4P_2(x)>0.
\]

Now let
\[
P_1(x)=24x^3+68x^2+53x+14.
\]
Its derivative is
\[
P_1'(x)=72x^2+136x+53,
\]
with roots
\[
x=-\frac{17}{18}\pm\frac{\sqrt{202}}{36}.
\]
Only
\[
x_1=-\frac{17}{18}+\frac{\sqrt{202}}{36}
\]
belongs to $[-1,1]$. At the endpoints and the critical point ($x_1$),
\[
P_1(-1)=5,\quad P_1(1)=159, \quad P_1(x_1)
=
\frac{2129-101\sqrt{202}}{486}>0.
\]
Therefore, $P_1(x)>0,~ x\in[-1,1].$
Since $1-x^2\geq0$ for $x\in[-1,1]$, we obtain
\[
8(1-x^2)P_1(x)\geq0.
\]
Finally, $48(1-x^2)^2(2x+1)^2\geq0.$

Thus, every coefficient on the right-hand side of \eqref{eq:a17} is non-negative. Since $t\geq0$, it follows that
\[
\frac{f(x,q)}{q^4}\geq0.
\]
As $q^4>0$, we conclude that
\[
f(x,q)\geq0,
\qquad x\in[-1,1],\quad q\in(0,1).
\]
Hence $B_1^2\beta_1 + \beta_5\rho^2 \geq 0.$ Similarly, $A_1^2\beta_2 + \beta_6\rho^2$ is non-negative (interchanging $q,m_1$ by $s,m_2$ respectively, in $B_1^2\beta_1 + \beta_5\rho^2$), and this implies $g'(\rho)$ is non-negative when $\rho \geq0$. 

Hence, if $\rho \geq0$, then $\rho^2_{1ss}$ and $\rho^2_{1cc}$ are both monotonically increasing in $\rho$, which implies $\rho^2_{1ss} + \rho^2_{1cc}$ is monotonically increasing in $\rho$. 

For $\rho < 0$, $f'(\rho)$ \eqref{eq:a14} and $g'(\rho)$ \eqref{eq:a15} are non-positive, and this implies $\rho^2_{1ss} + \rho^2_{1cc}$ is monotonic decreasing in $\rho$.

Now we consider $ \delta = -1$. Let $P_I(\cdot,\cdot)$ and $P_{II}(\cdot,\cdot)$ denote the pmfs for the cases $\delta = 1$ and $\delta = -1$, respectively. From the pmf given in \eqref{eq:2.5}, we obtain
\begin{equation*}
\resizebox{0.99\textwidth}{!}{$
\begin{aligned}
 P_{II}\bigg(\frac{2 \pi k}{m_1}, \frac{2\pi(2\beta-l)}{m_2}\bigg)& = C_1\big(q^{\zeta_1} + q^{m_1-\zeta_1}\big)\big(s^{\zeta_2} + s^{m_2-\zeta_2}\big)\bigg(1 + \rho \cos\bigg(\frac{2\pi \zeta_1}{m_1} - \frac{2\pi \zeta_2}{m_2}\bigg)\bigg)\\ 
 &= P_{I}\bigg(\frac{2 \pi k}{m_1}, \frac{2 \pi l}{m_2}\bigg), \quad \forall \hspace{0.1cm} (k,l) \in \mathbf{Z_{m_1}} \times\mathbf{Z_{m_2}}.     
\end{aligned}
$}
\end{equation*}
This establishes a one-to-one correspondence between the two cases. Thus, the result for the second case is a direct consequence of the first case.

\end{appendices}



\end{document}